\documentclass[reprint,superscriptaddress,amsmath,amssymb,aps,prb]{revtex4-2}

\usepackage[utf8]{inputenc}
\usepackage[T1]{fontenc}
\usepackage{graphicx}
\usepackage{dcolumn}
\usepackage{bm}
\usepackage{color}
\usepackage[colorlinks=true,linkcolor=blue,urlcolor=blue,citecolor=blue,pdfusetitle]{hyperref}
\usepackage{upgreek,textgreek}
\usepackage{physics}
\usepackage{orcidlink}
\usepackage{booktabs}

\begin{document}

\title{Mobile impurity interacting with a Hubbard chain and the role of Friedel oscillations}

\author{Felipe Isaule~\orcidlink{0000-0003-1810-0707}}
\affiliation{Instituto de Física, Pontificia
Universidad Católica de Chile,
Avenida Vicuña Mackenna 4860,
Santiago, Chile.}
\author{Abel Rojo-Franc\`{a}s~\orcidlink{0000-0002-0567-7139}}
\affiliation{Quantum Systems Unit, Okinawa Institute of Science and Technology Graduate University, Onna, Okinawa 904-0495, Japan.}
\affiliation{Departament de F{\'i}sica Qu{\`a}ntica i Astrof{\'i}sica, Facultat de F{\'i}sica, Universitat de Barcelona, E-08028 Barcelona, Spain}
\affiliation{Institut de Ci{\`e}ncies del Cosmos, Universitat de Barcelona, ICCUB, Mart{\'i} i Franqu{\`e}s 1, E-08028 Barcelona, Spain.}
\author{Duc Tuan Hoang~\orcidlink{0000-0002-8976-7157}}
\affiliation{Quantum Systems Unit, Okinawa Institute of Science and Technology Graduate University, Onna, Okinawa 904-0495, Japan.}
\author{Thom\'as Fogarty~\orcidlink{0000-0003-4940-5861}}
\affiliation{Quantum Systems Unit, Okinawa Institute of Science and Technology Graduate University, Onna, Okinawa 904-0495, Japan.}
\author{Thomas Busch~\orcidlink{0000-0003-0535-2833}}
\affiliation{Quantum Systems Unit, Okinawa Institute of Science and Technology Graduate University, Onna, Okinawa 904-0495, Japan.}
\author{Bruno Juliá-Díaz~\orcidlink{0000-0002-0145-6734}}
\affiliation{Departament de F{\'i}sica Qu{\`a}ntica i Astrof{\'i}sica, Facultat de F{\'i}sica, Universitat de Barcelona, E-08028 Barcelona, Spain}
\affiliation{Institut de Ci{\`e}ncies del Cosmos, Universitat de Barcelona, ICCUB, Mart{\'i} i Franqu{\`e}s 1, E-08028 Barcelona, Spain.}
%
\newcommand{\Hop}{\hat{H}}
\newcommand{\cop}{\hat{c}}
\newcommand{\copd}{\hat{c}^\dag}
\newcommand{\aop}{\hat{a}}
\newcommand{\aopd}{\hat{a}^\dag}
\newcommand{\dop}{\hat{d}}
\newcommand{\dopd}{\hat{d}^\dag}
\newcommand{\nop}{\hat{n}}
\newcommand{\UfI}{U_{fI}}
\newcommand{\UuI}{U_{\uparrow I}}
\newcommand{\UdI}{U_{\downarrow I}}
\newcommand{\argmax}{\text{argmax}}
\newcommand{\CfI}{\mathcal{C}_{fI}}
\newcommand{\ChI}{\mathcal{C}_{hI}}

\newcommand{\commentFelipe}[1]{\textbf{\textcolor{red}{[[FI: #1]]}}}
\newcommand{\changedFelipe}[1]{\textcolor{red}{#1}}

\date{\today}

\begin{abstract}

This work examines a mobile impurity interacting with a bath of a few spin-$\uparrow$ and spin-$\downarrow$ fermions in a small one-dimensional open lattice system. We study ground-state properties using the exact diagonalization method, where the system is modeled by a three-component Fermi Hubbard Hamiltonian. We find that in addition to the standard phase separation between a strongly repulsive impurity and the bath, a strongly-attractive impurity also phase separates with the fermionic holes due to the particle-hole symmetry. Furthermore, we find that the impurity can show an oscillatory pattern in its density for intermediate attractive and repulsive bath-impurity interactions, which are induced by Friedel oscillations in the finite-size fermionic bath. This rich behavior of the impurity could be probed with fermionic ultracold mixtures in optical lattices.

\end{abstract}

\maketitle

\section{Introduction}

The study of impurities interacting with a quantum bath is a fundamental problem in many-body physics. They often form dressed quasiparticles known as polarons, as coined in the seminal work of Landau and Pekar on electrons interacting with an ionic crystal~\cite{landau_effective_1948}. Since then, polarons have been extensively studied in many physical systems, such as in liquid Helium~\cite{bardeen_effective_1967}, quantum materials~\cite{alexandrov_polarons_2007,franchini_polarons_2021}, and even nuclear matter~\cite{tajima_intersections_2024}.

More recently, ultracold atomic mixtures~\cite{baroni_quantum_2024} have appeared as an ideal platform for realizing polaronic systems~\cite{massignan_polarons_2014,grusdt_impurities_2025,massignan_polarons_2025}. They offer a highly controllable setting where the interatomic interactions can be tuned by using Feshbach resonances~\cite{chin_feshbach_2010}, and impurities are realized by preparing mixtures with a high population imbalance. So far, these developments have enabled the experimental observation of both Fermi~\cite{schirotzek_observation_2009,nascimbene_collective_2009,kohstall_metastability_2012,koschorreck_attractive_2012,scazza_repulsive_2017} and Bose~\cite{hu_bose_2016,jorgensen_observation_2016,yan_bose_2020,skou_non-equilibrium_2021,henke_realization_2025}, polarons across their attractive and repulsive branches. In parallel to this, in the realm of 2D materials, it has also become possible to engineer impurities in transition metal dichalcogenides (TMDs)~\cite{mak_semiconductor_2022,massignan_polarons_2025}. In these, impurities can be realized with excitons interacting with electrons~\cite{chernikov_exciton_2014,courtade_charged_2017}, as pioneered by the observation of Fermi polarons in a MoSe$_2$ monolayer~\cite{sidler_fermi_2017}. Since then, further realizations of polarons in TMDs have emerged~\cite{tan_interacting_2020,tan_bose_2023}. 

This work focuses on impurities confined in a lattice system. This is motivated by the option of confining ultracold atoms in optical lattices~\cite{lewenstein_ultracold_2012,gross_quantum_2017} and of realizing Hubbard models with TMD bilayers~\cite{wu_hubbard_2018,tang_simulation_2020}. In this direction, impurities interacting with a bosonic bath in lattice systems have received extensive attention in the past few years~\cite{pasek_induced_2019,keiler_doping_2020,colussi_lattice_2023,ding_polarons_2023,santiago-garcia_lattice_2024,gomez-lozada_bosefermi_2025,isaule_bound_2024,isaule_counterflow_2025,alhyder_lattice_2025,hiyane_condensate-mediated_2025,hartweg_bose-hubbard_2025,christ_operator-valued-flow-equation_2025,dominguez-castro_polarons_2026,zhang_fate_2026}. Among these, topics of interest include the consideration of insulating baths~\cite{colussi_lattice_2023,isaule_counterflow_2025,alhyder_lattice_2025,zhang_fate_2026} and the onset of phase separation for strongly-repulsive impurities where the quasiparticle is destroyed~\cite{keiler_doping_2020,isaule_bound_2024,gomez-lozada_bosefermi_2025}. In turn, impurities immersed in a fermionic bath in the lattice have also been studied~\cite{amelio_polaron_2024-1,amelio_polaron_2024,hu_super_2024,hu_exact_2025,liu_exact_2025,pascual_polarons_2025}, especially for examining quasiparticle properties in two-dimensional geometries.

In this work, we examine the behavior of one mobile impurity interacting with a spin-$1/2$ balanced Fermi bath. We consider that all the particles are confined in a one-dimensional lattice within the tight-binding regime, which we model as a Hubbard Hamiltonian. Importantly, we consider a small open lattice which features Friedel oscillations in the fermionic bath~\cite{friedel_xiv_1952}. These correspond to a signature feature of fermions in the presence of defects. In our case, the oscillations are caused by the open boundary conditions, and thus, they are a finite-size effect, as reported in the past for Hubbard chains~\cite{bedurftig_friedel_1998}. 

We study the system using the exact diagonalization (ED) method~\cite{lin_exact_1993,sharma_organization_2015}, which has been employed successfully in the past for calculating Fermi polaron properties in two-dimensional lattices~\cite{amelio_polaron_2024-1}. We report an exhaustive examination of the system across a wide range of interactions, where it shows a very rich behavior despite its simplicity. On one side, we find that the system can show a phase separation between the impurity and fermionic holes for strong attraction. This is explained by the particle-hole symmetry of the Fermi Hubbard model~\cite{essler_one-dimensional_2005}, and thus it is not observed in the softcore bosonic lattice configurations studied in the past. On the other hand, we find that for intermediate bath-impurity interactions, the impurity displays peculiar patterns in its density, which are explained by the presence of Friedel oscillations in the bath. Thus, these findings suggest the use of impurities to probe Friedel physics in fermionic systems.

This work is organized as follows. Sec.~\ref{sec:model} presents the model, including the Hamiltonian and the numerical approach. Sec.~\ref{sec:1o4} presents the main results of this work, where the behavior of an impurity interacting with a fermionic bath at quarter filling is examined. Here, density profiles, correlations, and entanglement properties are examined. Then, from the understanding of the particle-hole symmetry of the model, Sec.~\ref{sec:3o4} generalizes the previous results to three quarters filling. Finally, Sec.~\ref{sec:conclusions} presents the conclusions and outlook of this work. Complementary results are also reported in the Appendix.

\section{Model}
\label{sec:model}

\subsection{Hamiltonian and setup}
\label{sec:model;sub:H}

We consider a one-dimensional optical lattice with $M$ sites loaded with a mobile impurity ($I$) and a bath of spin-$1/2$ fermions ($\sigma=\uparrow,\downarrow$). Such a three-component system can be achieved with ultracold atoms in different internal states, as done in the past with $^6$Li~\cite{ottenstein_collisional_2008,huckans_three-body_2009} (see Ref.~\cite{schumacher_observation_2026} for a recent realization). We assume that the particles interact through short-range on-site interactions and that the lattice has open boundary conditions. In the tight-binding regime, the system is modeled by a three-component Hubbard Hamiltonian
\begin{align}
    \Hop=&-t\sum_{\sigma=\uparrow,\downarrow} \sum_{i=1}^{M-1}\left(\copd_{i,\sigma}\cop_{i+1,\sigma}+\textrm{h.c.}\right)+U\sum_{i=1}^M \nop_{i,\downarrow}\nop_{i,\uparrow}\nonumber\\
    &-t \sum_{i=1}^{M-1}\left(\aopd_{i}\aop_{i+1}+\textrm{h.c.}\right)+\UfI\sum_{\sigma=\uparrow,\downarrow} \sum_{i=1}^M \nop_{i,\sigma}\nop_{i,I},
    \label{sec:model;sub:H;eq:H}    
\end{align}
where $\cop_{i,\sigma}$ ($\copd_{i,\sigma}$) and $\aop_i$ ($\aopd_i$) annihilate (create) a fermion of spin $\sigma$ and impurity at site $i$, respectively, and $\nop_{i,\sigma}=\copd_{i,\sigma}\cop_{i,\sigma}$ and $\nop_{i,I}=\aopd_i\aop_i$ are the respective number operators. 

The first two terms in Eq.~(\ref{sec:model;sub:H;eq:H}) describe the fermionic bath, the first corresponding to the tunneling term with amplitude $t>0$ and the second to the repulsive $\uparrow-\downarrow$ interaction with strength $U>0$. The third term describes the tunneling of the impurity, which has the same coupling $t$ as the fermions. Finally, the fourth term describes the fermion-impurity interaction, for which we consider equal (symmetric) interaction strengths $\UfI$ between the impurity and both spins. Additionally, this interaction can be either repulsive ($\UfI>0$) or attractive ($\UfI<0$). An examination of asymmetric interactions between the impurity and each spin state is presented in App.~\ref{app:asym}. There, we show that our main findings also hold for asymmetric interactions.

As mentioned, we consider one impurity, $N_I=1$, throughout. In turn, we consider a balanced bath with an equal number $N_f$ of fermions of each spin
\begin{equation}
    N_f=N_\uparrow=N_\downarrow.
    \label{sec:model;sub:H;eq:Nf}
\end{equation} 
Therefore, the filling factor is defined by
\begin{equation}
    \nu_f=N_f/M.
    \label{sec:model;sub:H;eq:nuf}
\end{equation}
This work focuses on baths with filling factors $\nu_f\neq 1/2$, which exhibit Friedel oscillations in open lattices~\cite{bedurftig_friedel_1998}. In the main text, we present results for the cases of $\nu_f=1/4$ and $\nu_f=3/4$, as these display the richest and most representative behavior. Other choices of filling factors are examined in Appendix~\ref{app:filling}.

\subsection{Exact diagonalization}
\label{sec:model;sub:ED}

To study the non-trivial effects that stem from the interacting impurity, we diagonalize the Hamiltonian using the ED method for a fixed number of particles~\cite{lin_exact_1993,sharma_organization_2015}.  We work with a mixed Hilbert space $\mathcal{H}=\mathcal{H}_\uparrow\otimes\mathcal{H}_\downarrow\otimes\mathcal{H}_I$ where the Fock states $|v_\sigma\rangle \in \mathcal{H}_\sigma$ correspond to distributions of the particles of species $\sigma=\uparrow,\downarrow,I$ across the lattice. These states read
\begin{equation}
    |v_\sigma\rangle=|n^{(v)}_{1,\sigma},...,n^{(v)}_{M,\sigma}\rangle,
\label{sec:model;sub:ED;eq:states}
\end{equation}
where $n^{(v)}_{i,\sigma}$ is the number of atoms $\sigma$ at site $i$ for a given state $v$, and thus $\sum_{i=1}^M n^{(v)}_{i,\sigma}=N_\sigma$. Additionally, because we work with either fermions or one impurity, $n^{(v)}_{i,\sigma}$ can only be 0 or 1. By using this Fock basis, a wavefunction reads
\begin{equation}
    |\Psi\rangle = \sum_{v_\uparrow = 1}^{\mathcal{D_\uparrow}}\sum_{v_\downarrow = 1}^{\mathcal{D_\downarrow}} \sum_{v_I = 1}^{\mathcal{D}_I} c_{v_\uparrow,v_\downarrow,v_I}|v_\uparrow\rangle|v_\downarrow\rangle|v_I\rangle,    
\end{equation}
where $\mathcal{D}_\sigma=\binom{M}{N_\sigma}$ is the size of the Hilbert space of each species $\mathcal{H}_\sigma$. Therefore, the size of the full Hilbert space is $\mathcal{D}=M\binom{M}{N_f}^2$. In turn, the coefficients $c_{v_\uparrow,v_\downarrow,v_I}$ are obtained from a numerical diagonalization of the Hamiltonian matrix.

Throughout this work, we focus on ground-state properties. From its wavefunction $|\Psi_0\rangle$, we calculate the density profiles
\begin{equation}
    n_{\sigma}(i)=\langle \nop_{i,\sigma}\rangle,\qquad n_I(i)=\langle \nop_{i,I}\rangle,
\label{sec:model;sub:ED;eq:nf}
\end{equation}
where $n_\sigma$ is the profile of each fermionic spin, while $n_I$ is that of the impurity. Note that, because in the main text the parameters for both spins are equal, we have that $n_\uparrow=n_\downarrow$ for all sites. 

In the following sections, most of our results consider $M=12$. However, an examination of other lattice sizes is reported in Appendix~\ref{app:size}.

\subsection{Particle-hole symmetry}
\label{sec:model;sub:PH}

An important aspect of the balanced [Eq.~(\ref{sec:model;sub:H;eq:Nf})] Hubbard model is its particle-hole symmetry~\cite{essler_one-dimensional_2005}. Importantly, this symmetry also imposes constraints on the three-component system considered in this work. As it will be shown in the next sections, the particle-hole symmetry explains the appearance of an impurity-hole phase separation for large $\UfI<0$, which is analogous to the standard impurity-fermion phase separation for large $\UfI>0$.

By performing a particle-hole transformation $\copd_{i,\sigma}\to \cop_{i,\sigma}$ in the Hamiltonian~(\ref{sec:model;sub:H;eq:H}) for $N_\uparrow=N_\downarrow$, one obtains that the density of each spin $\sigma=\uparrow,\downarrow$ satisfies
\begin{equation}
    n_\sigma(i;\UfI,\nu_f)+n_\sigma(i;-\UfI,1-\nu_f)=1.
    \label{sec:model;sub:PH;eq:nf}
\end{equation}
This relation simply exchanges particles by holes for opposite interaction strengths and filling factors. In turn, the density of the impurity satisfies
\begin{equation}
    n_I(i;\UfI,\nu_f)=n_I(i;-\UfI,1-\nu_f).
    \label{sec:model;sub:PH;eq:nI}
\end{equation}
These two relations impose important constraints on the densities of contrasting fillings, such as $\nu_f=1/4$ and $\nu_f=3/4$, which will be examined throughout this work. In the next section, we present a comprehensive examination of the system for $\nu_f=1/4$. Afterwards, in Sec.~\ref{sec:3o4}, we show analogous results for $\nu_f=3/4$, which derive easily from the understanding of the particle-hole symmetry relations. 

Finally, we stress that the particle-hole symmetry appears in the considered Hubbard model, which only holds in the tight binding approximation. To examine how the reported results change in shallower lattices, in Appendix~\ref{app:continous} we report complementary results for a continuous model with an oscillatory potential. 

\section{Quarter filling}
\label{sec:1o4}

In the following, we first examine the behavior of the fermions in the absence of the impurity to illustrate the onset of Friedel oscillations. Then, we report a comprehensive examination of the behavior of the interacting impurity.

\subsection{Friedel oscillations}
\label{sec:1o4;sub:Friedel}

\begin{figure}[t]
\centering
\includegraphics[width=\columnwidth]{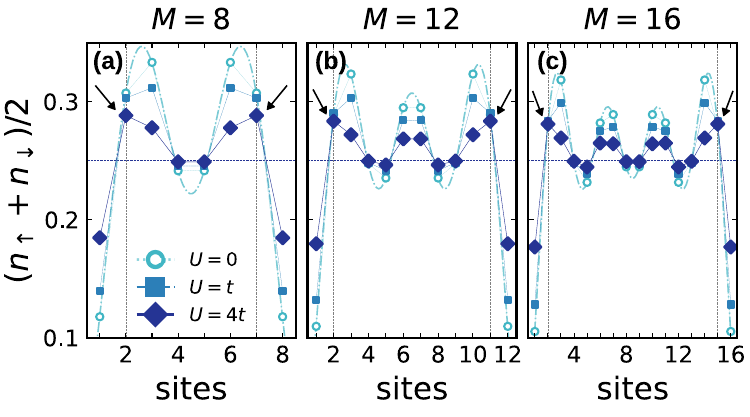}
\caption{Density profile of the fermions at $\nu_f=1/4$ in the absence of the impurity ($\UfI=0$) as a function of lattice sites. (a) considers a lattice with $M=8$, (b) with $M=12$, and (c) with $M=16$. The blank circles show results for $U=0$, the filled squares for $U=t$, and the filled diamonds for $U=4t$. The dashed horizontal lines indicate the filling factor of the fermions. The dash-dotted curves show the non-interacting profile from Eq.~(\ref{sec:1o4;sub:Friedel;eq:nonint}). }
\label{sec:1o4;sub:Friedel;fig:profiles}
\end{figure}

As mentioned, open finite lattices with $\nu_f\neq 1/2$ display an oscillatory pattern in the density $n_\sigma$ of the fermions. This corresponds to a border effect referred to as Friedel oscillations. To illustrate this for $\nu_f=1/4$, in Fig.~\ref{sec:1o4;sub:Friedel;fig:profiles}, we show a representative set of profiles of the fermions for $\UfI=0$. The figure displays results for different lattice sizes to illustrate the general nature of the profiles. The total density is divided by two to highlight the filling $\nu_f=1/4$. Additionally, we show the analytical result for the non-interacting limit, $U=0$. For $\nu_f<1/2$, this is given by~\cite{bedurftig_friedel_1998}
\begin{equation}
    n_\sigma(x)=\frac{N_\sigma+1/2}{M+1}-\frac{1}{2(M+1)}\frac{\sin\left(2\pi x \frac{N_\sigma+1/2}{M+1}\right)}{\sin\left(\frac{\pi x}{M+1}\right)},
    \label{sec:1o4;sub:Friedel;eq:nonint}
\end{equation}
for $\sigma=\uparrow,\downarrow$. The coordinate $x$ corresponds to a position in the lattice, with its integer values corresponding to specific sites $i=1,...,M$.

From Fig.~\ref{sec:1o4;sub:Friedel;fig:profiles}, it is easy to observe that all the panels exhibit the mentioned oscillatory pattern in the profiles. As shown by the panels, the number of peaks or maxima in the oscillations is equal to the number of fermions of each spin $N_f$, both in the non-interacting and interacting cases~\cite{bedurftig_friedel_1998}.  The latter is fulfilled for any filling factor $\nu_f<1/2$ with $N_f>1$.

A signature feature of Friedel oscillations, not only in lattices, is that the largest peaks are those closest to the boundaries. Indeed, the oscillations show a damping away from the borders. This behavior can be appreciated in the profiles for $M=12$ and $M=16$ [Figs.~\ref{sec:1o4;sub:Friedel;fig:profiles}(b) and~\ref{sec:1o4;sub:Friedel;fig:profiles}(c)], where the central oscillations have a smaller amplitude. Additionally, and very importantly for this work, we observe that the position of these largest maxima changes with $U$.

For small $U$ (see blank circles and filled squares), the largest peaks are located at the third site next to either border, $i=3$ and $i=M-2$~\cite{bedurftig_friedel_1998}. On the other hand, for large $U$ (see filled diamonds), the position of these largest peaks moves to the second site next to the border, $i=2$ and $i=M-1$, as indicated by the black arrows. This change of position is produced by the strong repulsion between fermions, which induces a spin-wave state. We have found that this change occurs around $U\approx 1.8 t$ for the examined lattice sizes. As we show in the following, the position of the peaks has an important effect on the behavior of the interacting impurity. 

Fig.~\ref{sec:1o4;sub:Friedel;fig:profiles} also shows that for large $U$ the oscillations become less pronounced, as they exhibit smaller deviations from the average density. While this might suggest that Friedel oscillations are less relevant at large repulsion, the following results show that the impurity is sensitive to these oscillations for a wide range of $U$. Additionally, the amplitude of the oscillations also decreases with increasing $M$. Therefore, the effect of Friedel oscillations examined in this work could become difficult to observe in very large lattices.

Finally, we note that the position of the peaks shows different behaviors for other filling factors. However, the qualitative impact of the Friedel oscillations on the impurity is general. We refer to Appendix~\ref{app:filling} for additional discussion on the filling factors.

\subsection{Densities with an interacting impurity}
\label{sec:1o4;sub:densities}

\begin{figure}[t]
\centering
\includegraphics[width=\columnwidth]{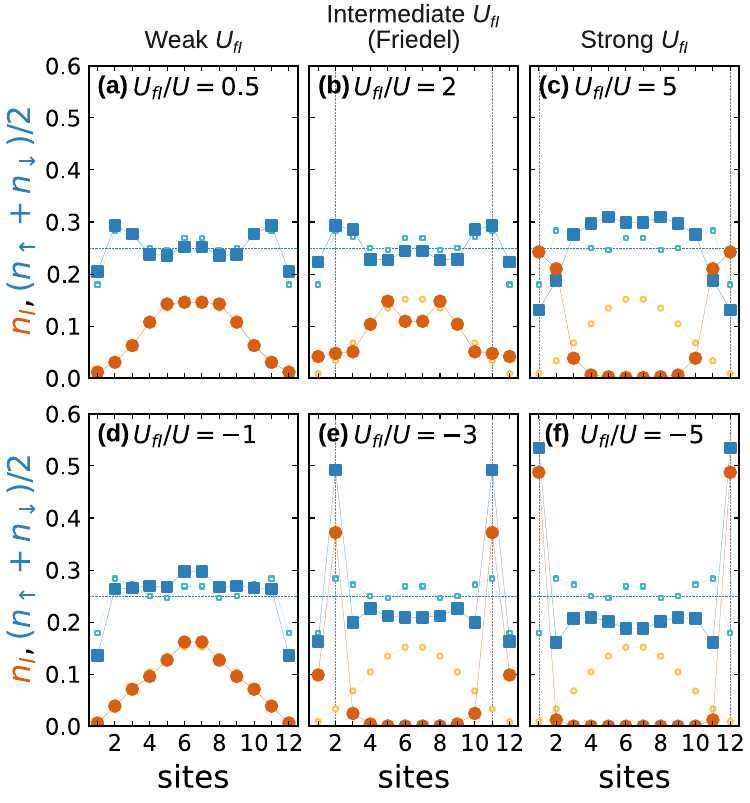}
\caption{Density profiles of the impurity (orange circles) and of the fermions (blue squares) at $\nu_f=1/4$ for $U=4t$ as a function of sites in a lattice with $M=12$. The filled large markers consider the finite interaction $\UfI$ indicated in each panel, while the blank small markers consider a non-interacting impurity ($\UfI=0$) for reference. The top panels and bottom panels consider repulsive and attractive interactions, respectively. The dashed horizontal lines indicate the filling factor.}
\label{sec:1o4;sub:densities;fig:densities}
\end{figure}

Having examined the bath's underlying behavior, we can now study the effect of the interacting impurity. In Fig.~\ref{sec:1o4;sub:densities;fig:densities} we show profiles of both the impurity and fermions for a representative set of fermion-impurity interactions $\UfI/U$. All the panels consider $U=4t$, with other values being examined later. In the following, we discuss different interaction regimes in turn.

\subsubsection{Weakly-interacting impurity and miscibility}

For small $|\UfI|/U$ [Figs.~\ref{sec:1o4;sub:densities;fig:densities}(a) and~\ref{sec:1o4;sub:densities;fig:densities}(d)], the impurity is weakly-interacting, and thus the overall behavior of the system is similar to that of the non-interacting limit $\UfI=0$. Here, the impurity (orange circles) shows a sine-like profile, with its peak at the center of the lattice. Note that, in both panels, the non-interacting profile (blank small orange circles) is almost completely underneath the larger markers for $\UfI\neq 0$. Similarly, the fermions (blue squares) show only a small deviation from their non-interacting profiles (blank small blue squares). 

Overall, from these profiles, we can infer that both the fermions and the impurity can occupy the same sites, with no important effect from the weak interaction $\UfI$. Therefore, we can identify these configurations as \emph{miscible} (M), consistent with the expected behavior for weakly-interacting impurities.

\subsubsection{Strongly-interacting impurity and phase separation}

We can now examine the regime of a strongly-interacting impurity. In the case of strong repulsion [Fig.~\ref{sec:1o4;sub:densities;fig:densities}(c)], we observe that the impurity localizes at the borders of the lattice, having a vanishing occupation at the center. In contrast, the fermionic bath occupies the central sites, decreasing its occupation at the borders. This behavior indicates a phase separation between the fermions and the impurity due to the strong repulsion, as is typically observed for strongly-repulsive impurities. Therefore, we refer to this configuration as a \emph{particle phase separation} (pPS). We note that the fermions do not exhibit a vanishing density at the borders due to the superposition of two states (phase separation at the right and the left). Nevertheless, we later show that fermions and impurities cannot occupy the same sites, confirming this immiscibility.

While the previous behavior can be expected, we observe a somewhat unexpected behavior for strong attraction [Fig.~\ref{sec:1o4;sub:densities;fig:densities}(f)]. Here, the impurity again localizes at the borders of the lattice. In turn, due to the strong bath-impurity attraction, the fermions also greatly increase their occupation at those bordering sites. Therefore, all the particles in the system show a large occupation at the borders. This can be counterintuitive, as one would expect that, for strong attraction, the system would collapse instead to the center of the lattice. As mentioned in~\ref{sec:model;sub:PH}, this new configuration also corresponds to a phase separation. However, this corresponds to a phase separation between the impurity and holes, instead of between the impurity and fermions (particles). Thus, we refer to this configuration as \emph{hole phase separation} (hPS), which stems from the particle-hole symmetry of the fermionic Hubbard chain. 

In general, in a phase separation, the minority species is pushed to the borders of the system. This explains why the impurity localizes at the borders for $|\UfI|\gg U$. In turn, the majority species occupies the center of the lattice. For strong repulsion, this majority species corresponds to the fermionic bath. On the other hand, for strong attraction, such a species corresponds to the holes. Thus, in the latter case, because the holes occupy the center of the lattice, the fermions must in turn increase their occupation at the borders, explaining the observed behavior in Fig.~\ref{sec:1o4;sub:densities;fig:densities}(f). 

One important difference between the two phase-separated configurations is the level of localization of the impurity.  In the observed pPS [Fig.~\ref{sec:1o4;sub:densities;fig:densities}(c)], the impurity can mostly occupy the last three sites at either border ($i=1,2,3$ and $i=M-2,M-1,M$). On the other hand, in the observed hPS [Fig.~\ref{sec:1o4;sub:densities;fig:densities}(e)], the impurity is sharply localized at sites $i=1$ and $i=M$. This is due to the low filling factor $\nu_f=1/4$. For pPS, the total number of fermions (particles) is smaller than the number of sites ($N_\uparrow+N_\downarrow<M$), leaving several free sites for the impurity to move to. In contrast, for hPS, the number of holes is greater than the number of sites. Therefore, in this hPS, it is energetically favorable to leave only a single site for the impurity at each border. 

Finally, we further stress that the hPS is a result of the particle-hole symmetry in the tight-binding Hubbard Hamiltonian. Therefore, the hPS configuration disappears in shallower lattices where such symmetry is not present. In contrast, the standard pPS is a general feature of strongly-repulsive systems, and thus it is still present in those shallower lattices. We refer to App.~\ref{app:continous} for a more detailed examination.

\subsubsection{Intermediate interactions and Friedel oscillations}

In addition to the previous configurations, we observe that for intermediate interactions, the impurity becomes influenced by the Friedel oscillations of the fermions.  This contrasts with the previously examined cases, where these oscillations do not play an important role. 

Firstly, for intermediate repulsion [Fig.~\ref{sec:1o4;sub:densities;fig:densities}(b)], we observe that the profile of the fermions shows only small deviations from that at $\UfI=0$,  continuing to show their $N_f=3$ peaks from their Friedel oscillations. Then, due to the fermion-impurity repulsion, the impurity prefers to localize between those peaks, where the fermions show a smaller occupation. This localization induces the impurity's profile to develop a two-peaks structure, deviating from its sine-like profile. Thus, $n_I$ displays an oscillatory pattern, which is due to the Friedel oscillations in the bath. Therefore, we refer to this configuration as \emph{Friedel repulsive} (FR).

Finally, for intermediate attraction [Fig.~\ref{sec:1o4;sub:densities;fig:densities}(e)], the impurity displays an even more striking localization at sites  $i=2$ and $i=M-1$, with a near vanishing occupation in the rest of the lattice. This localization is again a result of Friedel oscillations. Indeed, the underlying larger density of the fermions on sites $i=2$ and $i=M-1$, discussed in~\ref{sec:1o4;sub:Friedel}, induces the attractive impurity to localize to such sites. In turn, the impurity further enhances back the occupation of the fermions on those sites. In the following, we refer to this configuration as \emph{Friedel attractive} (FA).

Overall, the impurity displays oscillations in its density due to the presence of Friedel oscillations in the fermionic bath. As it will be shown in the following sections, the observed localizations of the impurity are also akin to the phase separations presented before. This means that the FR and FA regimes show features of pPS and hPS, respectively, being precursors of full phase separations.

\subsubsection{Dependence on the interaction strengths}

\begin{figure}[t]
\centering
\includegraphics[width=\columnwidth]{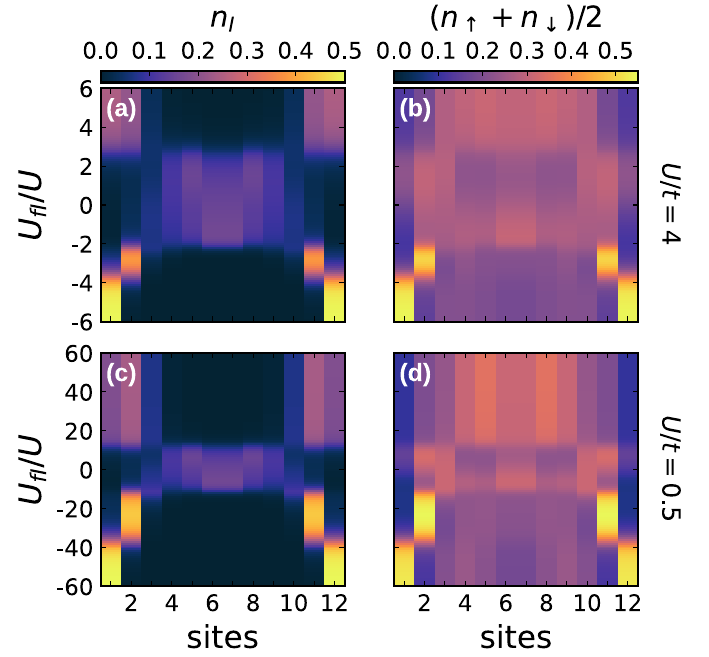}
\caption{Density profiles of the impurity [(a) and (c)] and of the fermions [(b) and (d)]  at $\nu_f=1/4$  as a function of $\UfI/U$ and of sites in a lattice with $M=12$. The top and bottom panels consider $U/t=4$ and $U/t=0.5$, respectively.}
\label{sec:1o4;sub:densities;fig:heatmaps}
\end{figure}

Having reported representative profiles of the different configurations, we now examine the dependence of the profiles on the interactions. In Fig.~\ref{sec:1o4;sub:densities;fig:heatmaps}, we report profiles of the impurity (left panels) and of the fermions (right panels) as a function of $\UfI$. The figure displays different density configurations, corresponding to the previously examined cases. Importantly, while the figure displays very well-defined regimes, we stress that the system shows crossovers instead of true phase transitions due to the consideration of small lattices and few particles.

For the previously examined case of $U/t=4$ (top panels), the pPS configuration occurs for $\UfI/U\gtrsim 3$, while the hPS one occurs for $\UfI/U \lesssim -4$. In turn, the Friedel-induced localization of the impurities at sites $i=2$ and $i=M-1$ (FA regime) occurs on a small but finite range of approximately $-4\lesssim \UfI/U\lesssim -2$. Importantly, within each configuration, the figure shows that the densities are approximately constant. Therefore, the profiles reported in Fig.~\ref{sec:1o4;sub:densities;fig:densities} are a good representation of each configuration.

One important aspect to examine is how the behavior of the system changes with $U/t$. We found that the same regimes appear for any $U/t>0$. However, the configurations appear for different ranges of $\UfI/U$. To illustrate this, we report densities for a small interaction of $U/t=0.5$ in the bottom panels of Fig.~\ref{sec:1o4;sub:densities;fig:heatmaps}. Overall, the densities show a similar behavior to that of the top panels, but with drastically different values of $\UfI$.

Interestingly, as discussed in~\ref{sec:1o4;sub:Friedel}, for small $U/t\lesssim 1.8$ and $\UfI=0$, the largest peaks in the fermionic density occur at sites $i=3$ and $i=M-2$. Thus, one would expect that for intermediate fermion-impurity attraction, the impurity would localize at those sites. However, the bottom panels of Fig.~\ref{sec:1o4;sub:densities;fig:heatmaps} show that, instead, the impurity still localizes at sites $i=2$ and $i=M-1$, as for larger values of $U$. In turn, due to the attraction, the peaks of the fermionic densities move to these sites. Therefore, for intermediate attraction $\UfI$, the impurity changes the pattern of the Friedel oscillations, with the system tending to develop peaks at sites  $i=2$ and $i=M-1$.

\subsection{Average position}
\label{sec:1o4;sub:x}

The previous results demonstrated that the system displays different configurations depending on the impurity's position. Therefore, one can characterize and identify these configurations by computing the average position of the impurity. Firstly, we calculate the site at which the impurity's profile $n_I$ is maximum, that is, the position of its peaks. Because the profiles are symmetric around the central site $i_c=(M+1)/2$, we compute this maximum from
\begin{equation}
    \text{site}^{(\text{max})}_I=|\argmax(n_I(i))|-i_c,
    \label{sec:1o4;sub:x;eq:sitemaxI}
\end{equation}
where $\argmax$ is the argument of the maximum of the function. At $\nu_f=1/4$, for an attractive impurity, we have found that this function can take the values
\begin{equation}
    \text{site}^{(\text{max})}_I(\UfI\leq 0)=\begin{cases}
    0 & :\text{w (M)}\\
    M/2-1/2 & : \text{sA (hPS)}\\
    \text{else} & : \text{FA}
    \end{cases},
    \label{sec:1o4;sub:x;eq:phasesA}
\end{equation}
allowing us to identify the weakly-interacting (w) miscible regime, the strongly-attractive (sA) regime with hPS, and the Friedel attractive (FA) regime
for intermediate interactions. On the other hand, for a repulsive impurity, it can take the values
\begin{equation}
    \text{site}^{(\text{max})}_I(\UfI\geq 0)=\begin{cases}
    0 & :\text{w (M)}\\
    M/2-1/2 & : \text{sR (pPS)}\\
    M/2-3/2 & : \text{sR (pPS)}\\
    \text{else} & : \text{FR}
    \end{cases},
    \label{sec:1o4;sub:x;eq:phasesR}
\end{equation}
which also allows us to identify the strongly-repulsive (sR) regime with pPS and the corresponding Friedel repulsive (FR) regime. Note that we found that the pPS configurations can appear for two values of $\text{site}^{(\text{max})}_I$. This is because, at $\nu_f=1/4$, the impurity has a large probability to occupy the last sites at each border [see Fig.~\ref{sec:1o4;sub:densities;fig:densities}(c)]. Thus, the maxima of $n_I$ can appear at any of those two sites.

In addition, we calculate a similarly defined average position of the impurity $\bar{x}_I$ as
\begin{equation}
    \bar{x}_I =\begin{cases}
    0 & :\argmax(n_I(i))=i_c\\
    \sum_{i=1}^M |i-i_c| n_I(i) & :\argmax(n_I(i))\neq i_c
    \end{cases}.
    \label{sec:1o4;sub:x;eq:xI}
\end{equation}
This average vanishes ($\bar{x}_I=0$) when the maximum of $n_I$ is at the central site, while it takes a finite value when the density shows non-centered peaks. While both functions give similar information, Eq.~(\ref{sec:1o4;sub:x;eq:sitemaxI}) is discrete and enables us to define interaction ranges for each configuration, which will be used as references during the rest of this article. However, as previously explained, our system displays crossovers. Therefore, the continuous function~(\ref{sec:1o4;sub:x;eq:xI}) better illustrates the crossover between the different regimes.

\begin{figure}[t]
\centering
\includegraphics[width=\columnwidth]{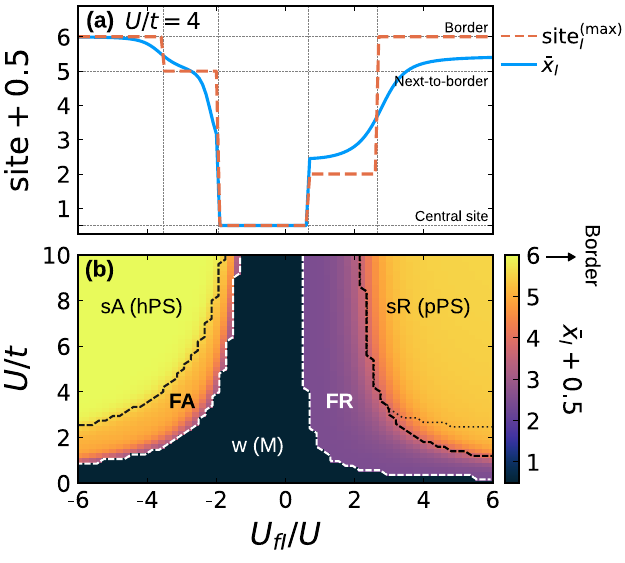}
\caption{(a) Average positions $\text{site}^{(\text{max})}_I$ (orange dashed line) and $\bar{x}_I$ (solid blue line) for $U/t=4$ as a function of $\UfI/U$. (b) Average position $\bar{x}_I$ as a function of $\UfI/U$ and of $U/t$. The vertical lines in (a) and dashed lines in (b) indicate changes in $\text{site}^{(\text{max})}_I$. The dotted line in (b) indicates a change of $\text{site}^{(\text{max})}_I$ within the same sR regime [see Eq.~(\ref{sec:1o4;sub:x;eq:phasesR})]. Both (a) and (b) consider $\nu_f=1/4$ and a lattice with $M=12$. A 0.5 is added to the average positions so that the sites are integer numbers.}
\label{sec:1o4;sub:x;fig:x}
\end{figure}

We show these average positions for the previously examined case of $U/t=4$ in Fig.~\ref{sec:1o4;sub:x;fig:x}(a). For comparison, the figure covers the same range of $\UfI/U$ reported in Fig.~\ref{sec:1o4;sub:densities;fig:heatmaps}. By following Eqs.~(\ref{sec:1o4;sub:x;eq:phasesA}) and (\ref{sec:1o4;sub:x;eq:phasesR}), we observe that the discrete average $\text{site}^{(\text{max})}_I$ (orange dashed line) enables us to define an interaction regime for each configuration. We indicate the separation between regimes with the vertical lines, which agree with the qualitative regimes that can be appreciated in Fig.~\ref{sec:1o4;sub:densities;fig:heatmaps}. On the other hand, $\bar{x}_I$ (solid blue line) shows a continuous behavior, as expected from a crossover. Nevertheless, $\bar{x}_I$ approximately follows the shape of $\text{site}^{(\text{max})}_I$, thus complementing its information. Additionally, in the strongly-repulsive regime [right side of Fig.~\ref{sec:1o4;sub:x;fig:x}(a)], the average $\bar{x}_I$ lies between the two last bordering sites. This correctly captures the behavior of the impurity at pPS, which can move within the last sites at either border.

Having demonstrated that the average positions can locate the different interaction regimes, in Fig.~\ref{sec:1o4;sub:x;fig:x}(b), we show the average $\bar{x}_I$ over a wide range of both $\UfI/U$ and $U/t$. Additionally, the dashed lines indicate the changes in $\text{site}^{(\text{max})}_I$ to locate each configuration regime. The figure shows well-defined regions covering a wide range of interactions, including the Friedel-induced regimes (FA and FR) for intermediate interactions. 

We find that the five configurations appear for any finite $U/t>0$. At very large $U/t$, the configuration regimes appear approximately within the same range of $\UfI/U$. This is illustrated by the dashed lines, which become approximately vertical lines for $U\gg t$. The latter might suggest a convergence to constant numbers. However, the Friedel configurations actually disappear in the limit $U/t\to\infty$, as this corresponds to the static limit where border effects, and so the Friedel oscillations, disappear. On the other hand, for very small $U/t$ [bottom region of Fig.~\ref{sec:1o4;sub:x;fig:x}(b)], the interactions at which each configuration appears change strongly with $U/t$. Indeed, the weak miscible regime persists for a wider range of $\UfI/U$, with the other regimes only appearing at stronger interactions. Importantly, as $U$ decreases, the transition interactions $|\UfI|/U$ between configurations increase non-linearly, showing the non-trivial effect of the tunneling. 

\subsection{Correlations}
\label{sec:1o4;sub:correlators}

\begin{figure}[t!]
\centering
\includegraphics[width=\columnwidth]{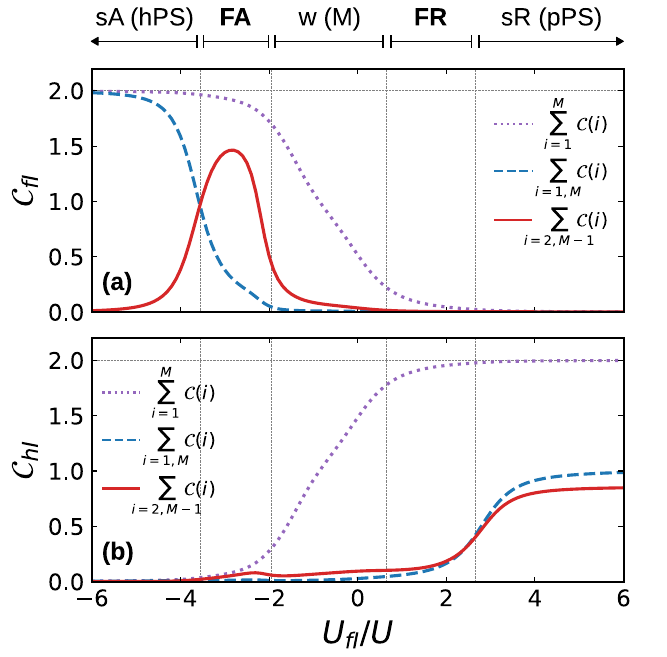}
\caption{Fermion-impurity (a) and hole-impurity (b) double occupation at $\nu_f=1/4$ for $U/t=4$ as a function of $\UfI/U$ in a lattice with $M=12$. The different lines represent the sum of double occupation across the sites indicated in the legends. The vertical lines indicate changes in $\text{site}^{(\text{max})}_I$, as reported in Fig.~\ref{sec:1o4;sub:x;fig:x}(a).}
\label{sec:1o4;sub:correlators;fig:C}
\end{figure}

\subsubsection{Fermion-impurity double occupation}
\label{sec:1o4;sub:correlators;sub:FI}

Having shown that the system supports different configurations depending on the fermion-impurity interaction $\UfI$, we can now further characterize the system by examining how the fermions and impurity correlate. We focus on two-body correlations of the form
\begin{equation}
    \CfI(i)=\sum_{\sigma=\uparrow,\downarrow}\langle \copd_{i,\sigma}\cop_{i,\sigma}\aopd_{i}\aop_{i}\rangle.
    \label{sec:1o4;sub:correlators;sub:FI;fig:C2FI}
\end{equation}
This corresponds to the fermion-impurity double occupation, which indicates how probable it is for the impurity and fermions to occupy the same site $i$. We show the double occupations $\CfI$ in Fig.~\ref{sec:1o4;sub:correlators;fig:C}(a). Additionally, the changes in the average occupation Eq.~(\ref{sec:1o4;sub:x;eq:sitemaxI}) are indicated by the vertical lines.

In the figure, the dotted purple line shows $\CfI$ summed across all sites in the lattice (total double occupation). A vanishing $\CfI$ for all sites means that the impurity and fermion never occupy the same sites. Therefore, we have that $\sum_{i=1}^M\CfI(i)=0$ for pPS. We observe in Fig.~\ref{sec:1o4;sub:correlators;fig:C}(a) that, indeed, this total $\CfI$ vanishes in the previously identified pPS regime. This confirms that such a regime corresponds to a phase separation between the fermions and the impurity. Furthermore, the figure shows that the total $\CfI$ is also small in the FR regime, showing that this regime starts showing phase-separated features, as previously mentioned. Importantly, we note that $\CfI$ shows a smooth decrease for increasing $\UfI>0$, and thus, there is no precise transition point to a phase-separated configuration.

On the other hand, we observe that the total $\CfI$ increases for decreasing $\UfI$, which is expected for attractive interactions where the double occupation becomes more favorable. In particular, we observe that the total double occupation reaches its maximum value of $\sum_{i=1}^M\CfI(i)=2$ for strong attraction. This maximum value indicates that the impurity always shares sites with both a spin-$\uparrow$ and a spin-$\downarrow$ fermion, akin to a trimer bound state. The figure shows that the total $\CfI$ converges to this value within the sA (hPS) regime, even though it also approximately reaches this value in the FA one. 

To further characterize the localization of the impurity at specific sites, the dashed blue line in Fig.~\ref{sec:1o4;sub:correlators;fig:C}(a) shows $\CfI$ summed only at the bordering sites $i=1$ and $i=M$. This quantity only becomes large in the sA regime with hPS, converging to $\sum_{i=1,M}\CfI(i)=2$ for strong attraction. Therefore, the total double occupation is completely dominated by these two bordering sites. Similarly, the solid red line in Fig.~\ref{sec:1o4;sub:correlators;fig:C}(a) shows $\CfI$ summed at the sites $i=2$ and $i=M-1$. This sum only becomes finite and large in the FA regime, which in the figure corresponds to $-4\lesssim \UfI/U \lesssim -2$. Because in this regime the impurity is sharply localized at $i=2$ and $i=M-1$, two fermions are also attracted to those sites. This further shows that the FA regime behaves similarly to a hPS, but with a localization at different sites. Nevertheless, note that the double occupation is not as strong as in the sA regime, reaching a maximum value of $\sum_{i=2,M-1}\CfI(i)\approx 1.5<2$.

Overall, the figure shows that $\CfI$ can be used to locate the different configurations. Therefore, the different sums of $\CfI$ complement the information provided by the average positions examined in~\ref{sec:1o4;sub:x}.

\subsubsection{Hole-impurity double occupation}
\label{sec:1o4;sub:correlators;sub:HI}

Additional information can be obtained for the double occupation between holes and the impurity. This is defined by
\begin{equation}
    \ChI(i)=\sum_{\sigma=\uparrow,\downarrow}\langle \cop_{i,\sigma}\copd_{i,\sigma}\aopd_{i}\aop_{i}\rangle,
    \label{sec:1o4;sub:correlators;sub:HI;fig:C2HI}
\end{equation}
where holes are created instead of fermions. By construction, the double occupations satisfy
\begin{equation}
    \sum_{i=1}^M\left(\CfI(i)+\ChI(i)\right)=2,
    \label{sec:1o4;sub:correlators;sub:HI;fig:Ctotal}
\end{equation}
showing that they are complementary quantities.

We display $\ChI$ in Fig.~\ref{sec:1o4;sub:correlators;fig:C}(b). In the figure, the dotted purple line shows that the total $\ChI$ increases for increasing $\UfI$, contrasting with $\CfI$. Naturally, this is a consequence of Eq.~(\ref{sec:1o4;sub:correlators;sub:HI;fig:Ctotal}). Thus, for strong repulsion where $\sum_{i=1}^M\ChI(i)=2$, the impurity and fermions phase separate, while the impurity forms a trimer-like state with two holes. In turn, the total $\ChI$ vanishes for strong attraction, confirming that a phase separation between the impurity and holes (hPS) occurs. Importantly, we observe that this total $\ChI$ is approximately zero in the FA regime, further showing that such a regime shows hPS features.

Fig.~\ref{sec:1o4;sub:correlators;fig:C}(b) additionally shows $\ChI$ summed at bordering sites (dashed blue and solid red lines). These do not dominate over different regimes as in Fig.~\ref{sec:1o4;sub:correlators;fig:C}(a). However, we observe that in the pPS regime $\sum_{i=1,M}\ChI(i)\approx\sum_{i=2,M-1}\ChI(i)\approx 1$, which is consistent with the impurity mostly occupying those sites [see Fig.~\ref{sec:1o4;sub:densities;fig:densities}(c)]. 

As will be explained in the following section, the hole-impurity double occupation is also important to explain the symmetric behavior for $\nu_f=3/4$.

\subsection{Entanglement}
\label{sec:1o4;sub:SvN}

Finally, we examine the von Neumann entropy to quantify the entanglement between the impurity and the fermionic bath. Importantly, this entropy can be used to probe transitions in many-body systems~\cite{amico_entanglement_2008}. In this direction, it has been proven useful to characterize multi-component systems~\cite{richaud_pathway_2019}, including the impurity limit~\cite{keiler_doping_2020}.

Here, we focus on the partial von Neumann entropy of the impurity. It reads
\begin{equation}
    S^{(\text{vN})}_{I} = -\tr\left[\rho_I\log\rho_I\right],
\end{equation}
where $\rho_I=\tr_f \rho$ is the reduced density matrix of the impurity, after tracing out the fermionic states $|v_\uparrow\rangle$ and $|v_\downarrow\rangle$ [see Eq.~(\ref{sec:model;sub:ED;eq:states})] from the full density matrix $\rho=|\Psi\rangle\langle \Psi|$. From a singular value decomposition, the entropy takes the form
\begin{equation}
    S^{(\text{vN})}_{I} = -\sum_{k=1}^M\lambda_k^{(I)}\log\lambda_k^{(I)},
\end{equation}
where $\lambda_k^{(I)}$ are the eigenvalues of $\rho_I$. Note that the sum has $M$ terms, as it corresponds to the size of the Hilbert space $\mathcal{D}_I$ of the impurity.

\begin{figure}[t!]
\centering
\includegraphics[width=\columnwidth]{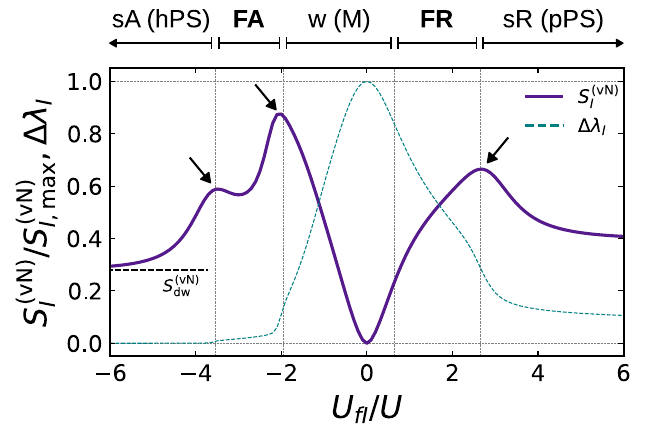}
\caption{Von Neumann entropy $S^{(\text{vN})}_{I}$ (solid line) and its Schmidt gap $\Delta \lambda_I$ (thin dashed line) at $\nu_f=1/4$ for $U/t=4$ as a function of $\UfI/U$ in a lattice with $M=12$. The vertical lines indicate changes in $\text{site}^{(\text{max})}_I$, as reported in Fig.~\ref{sec:1o4;sub:x;fig:x}(a). The short dashed line on the left indicates $S^{\text{(vN)}}_{\text{dw}}$ from Eq.~(\ref{sec:1o4;sub:SvN;eq:SvNdw}).}
\label{sec:1o4;sub:SvN;fig:SvN}
\end{figure}

We show the entropy for the previously examined case of $U/t=4$ in Fig.~\ref{sec:1o4;sub:SvN;fig:SvN}. The entropy is scaled by its possible maximum value
\begin{equation}
    S^{\text{(vN)}}_{I,\text{(max)}}=\log M,
    \label{sec:1o4;sub:SvN;eq:SvNmax}
\end{equation}
which is obtained from the condition of maximum entanglement, where $\lambda_k^{(I)}=1/M$ for all $k$. Additionally, the figure displays the Schmidt gap $\Delta \lambda_I$, which corresponds to the difference between the two largest eigenvalues $\lambda_k^{(I)}$.

From the figure, we observe that the entropy (solid line) is zero for $\UfI=0$, indicating that the impurity is not entangled with the fermions. Then, it initially increases as $|\UfI|$ increases within the weak and FR regimes. For attractive $\UfI$, the entropy reaches maximum values of approximately $0.8S^{\text{(vN)}}_{I,\text{(max)}}$, indicating a high, but not maximal, entanglement. In turn, it reaches maximum values of approximately $0.6S^{\text{(vN)}}_{I,\text{(max)}}$  for repulsive $\UfI$. Interestingly, after reaching these maxima, the entropy then decreases for stronger interactions. Thus, the system becomes less entangled, even though the interactions are stronger.

Importantly, while the entropy is a smooth function, which is expected from a system that displays crossovers, $S^{\text{(vN)}}_I$ shows abrupt changes between some regimes. Indeed, as indicated by the arrows, $S^{\text{(vN)}}_I$ reaches maxima approximately at some of the changes in $\text{site}^{(\text{max})}_I$ (vertical lines). Therefore, $\partial S^{\text{(vN)}}_I/\partial \UfI$ changes sign at these interactions. Here, we stress that the interaction regimes identified with the average position might be considered a heuristic approach. However, the changes in the von Neumann entropy around those points indicate that such crossovers do indeed correspond to meaningful physical changes in the system.

These physical changes correspond to the localization of the impurity. As explained in previous sections, we recall that the sA, sR, and even FA regimes can be interpreted as phase-separated configurations, where the impurity can only occupy a few sites. Therefore, the system is dominated by only a few Fock states [see Eq.~(\ref{sec:model;sub:ED;eq:states})], resulting in a decrease of $S^{\text{(vN)}}_I$. This is particularly evident deep in the sA regime, where the impurity is strongly localized at only two sites ($i=1$ and $i=M$), behaving like a particle in a balanced double well [see Fig.\ref{sec:1o4;sub:densities;fig:densities}(f)]. In this double well, we can take $M=2$ in Eq.~(\ref{sec:1o4;sub:SvN;eq:SvNmax}), obtaining
\begin{equation}
    S^{\text{(vN)}}_{\text{dw}}=\log 2.
    \label{sec:1o4;sub:SvN;eq:SvNdw}
\end{equation}
For $M=12$, this double-well entropy gives $S^{\text{(vN)}}_{I,\text{(dw)}}\approx 0.28 S^{\text{(vN)}}_{I,\text{(max)}}$, correctly predicting the saturation value for $\UfI\to-\infty$ shown in the left of Fig.~\ref{sec:1o4;sub:SvN;fig:SvN} (see dashed black line). On the other hand, in the FA and sR regimes, the impurity is not as strongly localized at only two sites, and thus we cannot easily obtain a similar double-well entropy. Nevertheless, the localization of the impurity explains the decrease of $S^{\text{(vN)}}_I$ in these regimes.

Following the previous arguments, we can also understand why the remaining crossover between the w and FR configurations does not show a change in $S^{\text{(vN)}}_I$. Instead, the entropy shows a monotonic increase with increasing $\UfI$ between such regimes. This is because in the FR regime, while it shows oscillations in $n_I$, the impurity is still delocalized across the lattice [see Fig.~\ref{sec:1o4;sub:densities;fig:densities}(b)], as in the miscible weak regime. 

Finally, the Schmidt gap (thin dashed line) is, by construction, one in the limit of no entanglement ($\UfI=0$). Then, it shows a monotonic decrease with increasing $|\UfI|$ in all regimes, without great changes at the transitions. Nevertheless, we note that $\Delta \lambda_I$ vanishes for large $|\UfI|$, where the ground state becomes degenerate.

\section{Three quarters filling}
\label{sec:3o4}

We have shown that at quarter filling $\nu_f=1/4$ the system shows a rich behavior. It features phase separations between the impurity and either the fermions or the holes, and also configurations where the position of the impurity is influenced by the Friedel oscillations of the bath. Overall, the system displays analogous configurations for different fillings, even though the precise behavior depends on the specific filling factor.

In this section, we briefly examine the case of $\nu_f=3/4$, building upon the results from the previous section. A complementary examination of other filling factors is provided in Appendix~\ref{app:filling}.

\subsection{Friedel oscillations}
\label{sec:3o4;sub:Friedel}

\begin{figure}[t]
\centering
\includegraphics[width=\columnwidth]{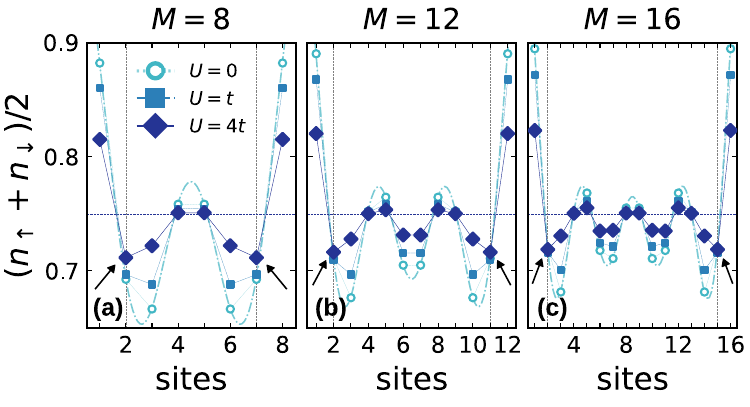}
\caption{Density profile of the fermions at $\nu_f=3/4$ in the absence of the impurity ($\UfI=0$) as a function of sites. (a) considers a lattice with $M=8$, (b) with $M=12$, and (c) with $M=16$. The blank circles show results for $U=0$, the filled squares for $U=t$, and the filled diamonds for $U=4t$. The dashed horizontal lines indicate the filling factor of the fermions.  The dash-dotted curves show the non-interacting profile from Eq.~(\ref{sec:3o4;sub:Friedel;eq:nonint}).}
\label{sec:3o4;sub:Friedel;fig:profiles}
\end{figure}

We start again by examining the onset of Friedel oscillations of the fermions for $\UfI=0$. Open lattices with $\nu_f \geq 1/2$ also display these oscillations in the densities of the fermions. However, they show contrasting occupations, as dictated by the particle-hole symmetry condition~(\ref{sec:model;sub:PH;eq:nf}). We illustrate this in Fig.~\ref{sec:3o4;sub:Friedel;fig:profiles}, where we show densities for $\nu_f=3/4$.

The figure exhibits the same oscillatory pattern displayed previously in Fig.~\ref{sec:1o4;sub:Friedel;fig:profiles} for $\nu_f=1/4$. However, the oscillations are clearly mirrored in the $y$-axis, following Eq.~(\ref{sec:model;sub:PH;eq:nf}). Thus, by using Eq.~(\ref{sec:1o4;sub:Friedel;eq:nonint}), the analytical expression for $U=0$ and $\nu_f>1/2$ takes the form
\begin{equation}
    n_\sigma(x)=\frac{N_{h,\sigma}+1/2}{M+1}+\frac{1}{2(M+1)}\frac{\sin\left(2\pi x \frac{N_f+1/2}{M+1}\right)}{\sin\left(\frac{\pi x}{M+1}\right)},
    \label{sec:3o4;sub:Friedel;eq:nonint}
\end{equation}
where $N_{h,\sigma}=M-N_\sigma$ is the number of holes.

Due to the symmetry, the peaks in the occupations at $\nu_f=1/4$ are now replaced by local minima at $\nu_f=3/4$. Importantly, these minima follow the same properties shown previously by the maxima. Thus, the largest minima are those closest to the border of the lattice. We observe that these largest minima also move from the third site next to either border at small $U/t$, to the second sites $i=2$ and $i=M-1$ for larger $U/t$ (indicated by the black arrows). More generally, the number of these minima is equal to the number of holes of each spin $N_{h,\sigma}$. This is fulfilled for any filling factor $\nu_f>1/2$ with $N_f<M-1$.

\subsection{Interacting impurity}
\label{sec:3o4;sub:Impurity}

\begin{figure}[t]
\centering
\includegraphics[width=\columnwidth]{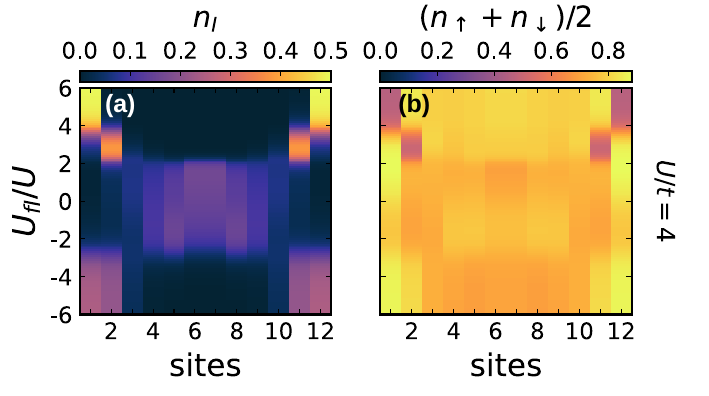}
\caption{Density profile of the impurity (a) and of the fermions (b) at $\nu_f=3/4$ for $U/t=4$ as a function of $\UfI/U$ and of sites in a lattice with $M=12$. }
\label{sec:3o4;sub:Impurity;fig:heatmaps}
\end{figure}

We can now examine the behavior of the system under the influence of the interacting impurity. We show densities of both the impurity and the fermions for varying $\UfI$ in Fig.~\ref{sec:3o4;sub:Impurity;fig:heatmaps}, which is analogous to Figs.~\ref{sec:1o4;sub:densities;fig:heatmaps}(a)-(b). The impurity [Fig.~\ref{sec:1o4;sub:densities;fig:heatmaps}(a)] shows an identical profile to that of Figs.~\ref{sec:1o4;sub:densities;fig:heatmaps}(a) after exchanging $\UfI\to -\UfI$. This follows the condition~(\ref{sec:model;sub:PH;eq:nI}) from the particle-hole symmetry. In turn, the fermions exhibit an analogous profile that follows Eq.~(\ref{sec:model;sub:PH;eq:nf}). 

For very strong repulsion $\UfI\gtrsim 4U$, the system undergoes a pPS where the impurity is sharply localized at sites $i=1$ and $i=M$, while the fermions decrease their densities at those sites due to the strong repulsion. Therefore, this corresponds to a phase separation between the fermions and the impurity. This pPS regime at $\nu_f=3/4$ is thus symmetric to the hPS at $\nu_f=1/4$ for the opposite values of $\UfI$ due to the particle-hole symmetry. Here we note that while the hPS for $\nu_f=1/4$ was somewhat counterintuitive, the pPS is the expected behavior for strong repulsion. Therefore, Fig.~\ref{sec:3o4;sub:Impurity;fig:heatmaps} further illustrates the immiscible nature of both the pPS and hPS phases.

Then, at intermediate repulsion, in Fig.~\ref{sec:3o4;sub:Impurity;fig:heatmaps}(a), we observe a sharp localization of the impurity at sites $i=2$ and $i=M-1$. Conversely, this corresponds to an FR regime with pPS properties, being symmetric to the FA regime with hPS properties at $\nu_f=1/4$. From this, it becomes easier to conclude that for attractive interactions we observe FA and hPS regimes at $\nu_f=3/4$, which are symmetric to the FR and pPS at $\nu_f=1/4$.

\begin{figure}[t!]
\centering
\includegraphics[width=\columnwidth]{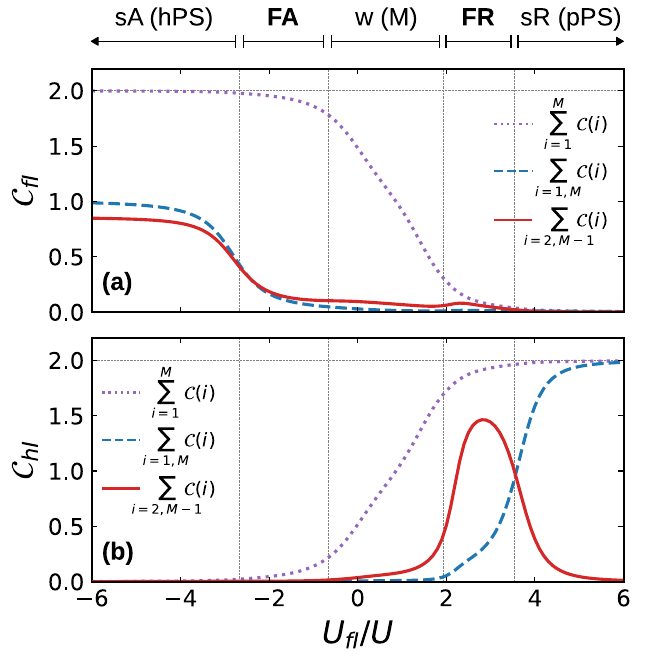}
\caption{Fermion-impurity (a) and hole-impurity (b) double occupation at $\nu_f=3/4$ for $U/t=4$ as a function of $\UfI/U$ in a lattice with $M=12$. The different lines represent the sum of double occupation across the sites indicated in the legends.  The vertical lines indicate changes in $\text{site}^{(\text{max})}_I$.}
\label{sec:3o4;sub:Impurity;fig:C}
\end{figure}

Overall, we can report that the behaviors of the system at one and three-quarters fillings are completely symmetric. Therefore, we do not repeat the examination of the average occupations and entanglement, as the results are identical to those at $\nu_f=1/4$ after exchanging the sign of $\UfI\to-\UfI$. However, it is worth examining the double occupations again. These are shown for $\nu_f=3/4$ in Fig.~\ref{sec:3o4;sub:Impurity;fig:C}. From the figure, one can observe that the double occupations are identical to those at $\nu_f=1/4$ (Fig.~\ref{sec:1o4;sub:correlators;fig:C}) after exchanging $\UfI\to-\UfI$, but also the correlators $\CfI\to\ChI$ and $\ChI\to\CfI$. This further illustrates the particle-hole symmetric behavior of the system, where the correlations between impurity and fermions are replaced by symmetric correlations between impurity and holes, and vice versa.

\section{Conclusions}
\label{sec:conclusions}

In this work, we reported a comprehensive study of a mobile impurity interacting with a spin-$1/2$ fermionic bath in small one-dimensional lattices. We considered finite open lattices, which give rise to Friedel oscillations in the fermionic bath. We found that the system shows rich behavior across different bath-impurity interactions. The system displays different configurations, including a miscible regime, phase separation between the impurity and either fermions or their holes, and intermediate configurations where the bath's Friedel oscillations influence the impurity.

The hole-impurity phase separation produces, counterintuitively, an increase in the density of all particles in the borders of the lattice for strong fermion-impurity attraction. This contrasts with the expected collapse to the center of the lattice, as in weakly-interacting bosonic systems. We have shown that this particular behavior is due to the particle-hole symmetry of the employed fermionic Hubbard model. On the other hand, for intermediate fermion-impurity interactions, the impurity is influenced by the Friedel oscillations of the bath. Here, the impurity develops non-trivial patterns and localizations in its density.

We characterized the different configurations by examining the average position of the impurity in the lattice and two-body correlations. Additionally, we examined the entanglement between species. We found that the partial von Neumann entropy changes its behavior around the crossovers between regimes, proving that each configuration changes the correlations between the fermions and the impurity.

The found results highlight features displayed by impurities interacting with fermions in lattices, not present in the bosonic systems widely studied in the past few years. Furthermore, our results suggest the use of impurities to probe Friedel oscillations in fermionic systems. Extensions of this work include the study of dynamics and, more importantly, the exploration of Friedel physics in non-lattice configurations.

\begin{acknowledgments}
 We thank J. Martorell for useful discussions. F.I. acknowledges funding from ANID through FONDECYT Postdoctorado No. 3230023. 
 This work has been funded by Grant PID2023-147475NB-I00 funded by MICIU/AEI/10.13039/501100011033 and FEDER, UE, by grants 2021SGR01095 from Generalitat de Catalunya, and by Project CEX2024-001451-M of ICCUB (Unidad de Excelencia María de Maeztu). 
This work was supported by the Okinawa Institute of Science and Technology Graduate University. D.T.H. is grateful for the Scientific Computing and Data Analysis (SCDA) section of the Research Support Division at OIST. T.F. acknowledges support from JSPS KAKENHI Grant No. JP23K03290. T.F. and T.B. are also supported by JST Grant No. JPMJPF2221.
\end{acknowledgments}

\appendix

\section{Asymmetric fermion-impurity interactions}
\label{app:asym}

The presented results focus on the specific case where both fermionic spins interact with the impurity via the same interaction strength $\UfI$. However, in an experiment, each spin could interact differently with the impurity. For example, we refer to Ref.~\cite{huckans_three-body_2009,schumacher_observation_2026} for scattering lengths between different $^6$Li spin states. Nevertheless, in those experiments, the magnetic field could be chosen to have near-symmetric fermion-impurity interactions depending on the chosen impurity spin state.

To examine how the presented results hold for asymmetric interactions, here we consider the following bath-impurity interaction Hamiltonian term
\begin{equation}
    \Hop_{fI}=\UuI \sum_{i=1}^M \nop_{i,\uparrow}\nop_{i,I}+\UdI \sum_{i=1}^M \nop_{i,\downarrow}\nop_{i,I},
\end{equation}
where $\UuI$ and $\UdI$ are the interaction strengths between the impurity and the spin-$\uparrow$ and spin-$\downarrow$ fermions, respectively. The symmetric configuration examined in the main text is then recovered for $\UfI=\UuI=\UdI$.

\begin{figure}[t]
\centering
\includegraphics[width=\columnwidth]{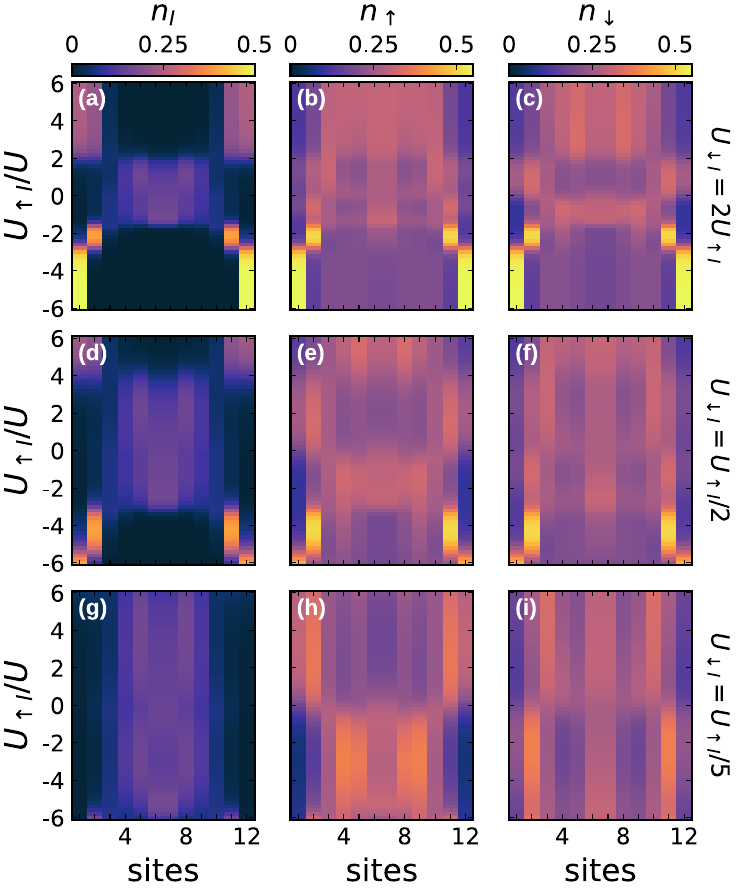}
\caption{Density profiles of the impurity [(a), (d), and (g)], of the spin-$\uparrow$ fermions [(b), (e), and (h)], and of the spin-$\downarrow$ fermions [(c), (f), and (i)] for $U/t=4$ as a function of $\UuI/U$ and of sites in a lattice with $M=12$ sites. (a)--(c) consider $\UdI=2\UuI$, (d)--(f) consider $\UdI=\UuI/2$, and (g)--(i) consider $\UdI=\UuI/5$.}
\label{app:asym;fig:profiles}
\end{figure}

We show examples of density profiles for asymmetric interactions in Fig.~\ref{app:asym;fig:profiles}. In the figure, we vary the interaction strength $\UuI$, while keeping the ratio $\UdI/\UuI$ fixed for the panels in the same row. Due to the asymmetry, we have that $n_\uparrow\neq n_\downarrow$, and thus we display the densities of each spin separately. 

From the figure, we observe that for $\UdI=2\UuI$ (upper panels) and $\UdI=\UuI/2$ (middle panels), the system behaves similarly to the symmetric case (see Fig.~\ref{sec:1o4;sub:densities;fig:heatmaps}). Indeed, the system displays the same configurations. For example, the panels clearly show that around $\UuI/U\approx -4$, the FA configuration where the impurity sharply localizes at the second site near each border also appears, attracting both spins to such sites. The asymmetry in the interactions only makes the regimes appear within slightly different interaction strengths and induces small differences in the densities. Therefore, our results hold for small asymmetries in the interactions.

The situation changes slightly for large asymmetries. For example, the bottom panels of Fig.~\ref{app:asym;fig:profiles} consider $\UdI=\UuI/5$. For this large asymmetry, the system does not exhibit the same configurations as those examined before for the shown range of $\UuI$. For instance, there is no FA regime in which all the particles increase their densities at sites $i=2$ and $i=M-1$. However, the effect of Friedel oscillations is still present. At around $\UuI/t \approx -4$, panels (g) and (h) show that both the impurity and spin-$\uparrow$ fermions increase their density around sites $i=4$ and $i=8$ due to their attraction. Thus, we refer to this regime as FA$\downarrow$. In contrast, panel (i) shows that the spin-$\downarrow$ fermions increase their density at sites $i=2$ and $i=M-1$, being somewhat separated from the impurity. Nevertheless, we stress that additional regimes could appear at larger values of $|\UuI|$.

\begin{figure}[t]
\centering
\includegraphics[width=\columnwidth]{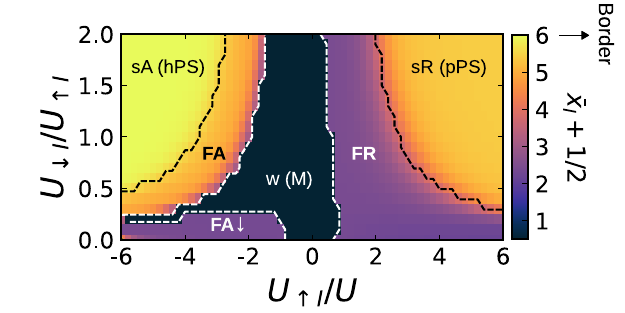}
\caption{Average position $\bar{x}_I$ as a function of $\UuI/U$ and of $\UdI/t$ for $U/t=4$. A 0.5 is added to the average positions so that the sites are integer numbers.}
\label{app:asym;fig:x}
\end{figure}

To further show the dependence on the asymmetry, in Fig.~\ref{app:asym;fig:x} we show the average positions as a function of the interactions to illustrate the different regimes. We observe that, for $\UdI/\UuI \gtrsim 0.3$, the system shows the configurations presented in the main text, displaying a qualitatively similar behavior to that for $\UdI=\UuI$. However, for $\UdI/\UuI \lesssim 0.3$ we observe a different behavior, with the onset of the FA$\downarrow$ regime discussed before. Additionally, we note that while we have referred to the regime of intermediate and large repulsive $\UuI$ as FR, both spins display different Friedel patterns, as shown in the upper halves of Fig.~\ref{app:asym;fig:profiles}(h) and Fig.~\ref{app:asym;fig:profiles}(i). However, a complete examination of the system under such high asymmetries is left for future work.

Finally, we mention that we also examined asymmetric interactions with opposite sign, which are relevant for magnetic impurities. We found that phase-separated and Friedel regimes also emerge in such cases. However, and as expected, the system shows a much more complex behavior, with miscible phases with only one spin, among other configurations.

In summary, the results reported in the main text are robust for small asymmetries in the interaction. Additionally, while larger asymmetries give rise to additional configurations, phase separations with holes or particles and regimes affected by Friedel oscillations are still present. Thus, we can conclude that the main conclusions of this work hold for asymmetric interactions. 

\section{Dependence on the filling factor}
\label{app:filling}

\begin{figure}[t]
\centering
\includegraphics[width=\columnwidth]{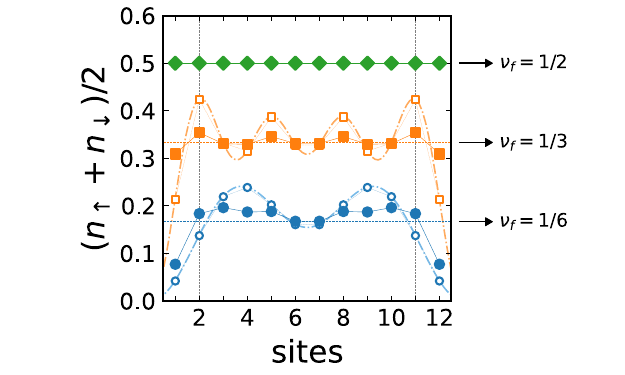}
\caption{Density profile of the fermions at $\nu_f=1/2$ (green diamond), $\nu_f=1/3$ (orange squares), and $\nu_f=1/6$ (blue circles) in the absence of the impurity ($\UfI=0$) as a function of sites in a lattice with $M=12$. The blank markers show results for $U=0$, while the filled markers for $U=4t$. The dashed horizontal lines indicate the filling factor of the fermions. The dash-dotted curves show the non-interacting profile from Eq.~(\ref{sec:1o4;sub:Friedel;eq:nonint}) for $\nu_f=1/3$ and $\nu_f=1/6$. }
\label{app:filling;fig:Friedel}
\end{figure}

In this appendix, we examine how the reported results change for other filling factors and lattice sizes. Firstly, in Fig.~\ref{app:filling;fig:Friedel} we display densities of the fermions for $\UfI=0$ at different filling factors $\nu_f\leq 1/2$. We do not show profiles for $\nu_f>1/2$, as they are simply symmetric to those for $\nu_f < 1/2$, as dictated by Eq.~(\ref{sec:model;sub:PH;eq:nf}).

At half filling $\nu_f=1/2$ (green diamonds), the profiles are constant, $n_\sigma(i)=1/2$ for all $i$, for any interaction $U>0$. These simply correspond to insulating configurations, as mentioned in the main text. In contrast, lower filling factors exhibit Friedel oscillations, with a number of maxima equal to the number of fermions of each spin. Indeed, for $\nu_f=1/3$ (orange squares), the oscillations show four peaks, consistent with having four fermions of each spin for $M=12$ sites. Similarly, for $\nu_f=1/6$ (blue circles), the oscillations show two peaks, as the system has only two fermions of each spin.  Importantly, the positions of the largest peaks show a dependence on the filling factor. At $\nu_f=1/3$, the largest peaks are always at sites $i=2$ and $i=M-1$, independently of $U$. In contrast, at $\nu_f=1/6$, the positions of the largest peaks do change with the interaction, also becoming less pronounced with increasing $U$.

\begin{figure}[t]
\centering
\includegraphics[width=\columnwidth]{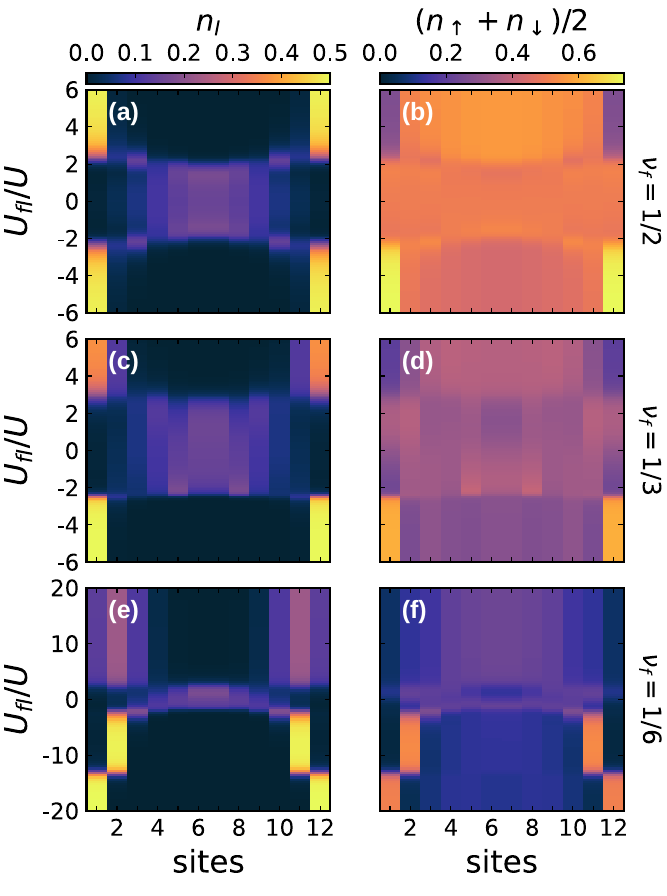}
\caption{Density profiles of the impurity [(a), (c), and (e)] and of the fermions [(b), (d), and (f)] for $U/t=4$ as a function of $\UfI/U$ and of sites in a lattice with $M=12$ sites. (a) and (b) consider $\nu_f=1/2$, (c) and (d) consider $\nu_f=1/3$, and (e) and (f) consider $\nu_f=1/6$.}
\label{app:filling;fig:Heatmaps}
\end{figure}

The effect of the interacting impurity is then illustrated in Fig.~\ref{app:filling;fig:Heatmaps}, which shows densities of the impurity (left panels) and of the fermions (right panels) for varying $\UfI$ at different filling factors. At half filling (top panels), the system shows a weak miscible regime for small $|\UfI|$, corresponding to the region $-2 \lesssim \UfI/U \lesssim 2$ in the figure. Then, for strong repulsion (upper regions in the top panels), the system expectedly phase-separates. The impurity becomes sharply localized at the borders, while the fermions gather at the center of the system. Therefore, such a regime corresponds to a pPS. Similarly, for strong attraction (bottom regions), the impurity also localizes at the same bordering sites, attracting the fermions. We can again identify such a regime as an hPS. Note that at half-filling, condition~(\ref{sec:model;sub:PH;eq:nI}) becomes $n_I(\UfI)=n_I(-\UfI)$, and thus the impurity exhibits a completely symmetric profile around $\UfI$. Most importantly, at $\nu_f=1/2$ there is no Friedel regime, as Friedel oscillations are not present at this filling. 

However, regimes influenced by Friedel oscillations appear for any other filling factor. At $\nu_f=1/3$ [Figs.~\ref{app:filling;fig:Heatmaps}(c)-(d)], the system shows the expected pPS and hPS for strong repulsion and attraction, respectively (upper and bottom parts of the panels). At weaker interactions, the impurity shows oscillatory patterns, as can be appreciated for $-2\lesssim \UfI/U \lesssim 2$ in Fig.~\ref{app:filling;fig:Heatmaps}(c). While the oscillations are small and the impurity does not exhibit any sharp localization at particular sites, these patterns are caused by Friedel oscillations. Therefore, we can identify those configurations as Friedel regimes.

Finally, the bottom panels of Fig.~\ref{app:filling;fig:Heatmaps} show similar densities to those observed for $\nu_f=1/4$. Indeed, we observe near-identical phase-separated and Friedel regimes, including a sharp localization of the impurity at sites $i=2$ and $i=M-1$ around $\UfI/U\approx -10$. Nonetheless, note that the regimes occur at much larger interactions $\UfI/U$, as can be appreciated in the change of the $y$-axis numbers.

In summary, the system shows pPS and hPS configurations for any filling factor, and also shows Friedel regimes for $\nu_f\neq 1/2$. The Friedel regimes manifest as oscillatory patterns in the density of the impurity, which develops non-centered peaks away from the borders. However, the particular form of the impurity's profile depends on the specific filling factor.

\section{Dependence on the lattice size}
\label{app:size}

.

\begin{figure}[t]
\centering
\includegraphics[width=\columnwidth]{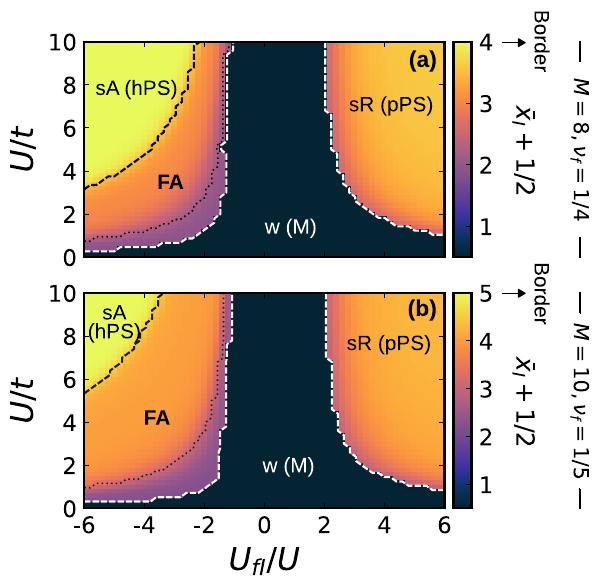}
\caption{Average position $\bar{x}_I$ as a function of $\UfI/U$ and of $U/t$. The dashed lines indicate changes in $\text{site}^{(\text{max})}_I$.  (a) considers $\nu_f=1/4$ and a lattice with $M=8$, while (b) consider $\nu_f=1/5$ and a lattice with $M=10$. The dotted lines indicate a change of $\text{site}^{(\text{max})}_I$ within the same FA regime.  A 0.5 is added to the average positions so that the sites are integer numbers.}
\label{app:size;fig:x}
\end{figure}

The previous appendix showed that phase-separated configurations and Friedel regimes appear for different filling factors with $M=12$ sites. We have checked that we obtain similar results for other lattice sizes, obtaining analogous profiles. To illustrate that the different regimes appear for other numbers of sites, in Fig.~\ref{app:size;fig:x} we show the average positions over a wide range of interactions for both $M=8$ [panel (a)] and $M=10$ [panel (b)]. This figure is analogous to Fig.~\ref{sec:1o4;sub:x;fig:x}(b). For $M=8$, we consider the same quarter filling. However, for $M=10$ we consider a filling of $\nu_f=1/5$, as $\nu_f=1/4$ is not possible for ten sites.

Overall, both panels show a qualitatively similar behavior as that reported for $M=12$, further illustrating the general nature of our results. The interaction ranges at which each configuration appears show some dependence on $M$, which can be expected. Importantly, these lattice sizes do not support FR configurations. This is because, at these filling factors, the lattice sizes only have two fermions of each spin. Therefore, the bath does not have the two non-centered minima required to induce a pattern in $n_I$ for intermediate repulsion.

We also note that the FA regime has regions where $\text{site}^{(\text{max})}_I=M-1$ (orange region) and where $\text{site}^{(\text{max})}_I=M-2$ (purple region), which are separated by the dotted lines. The first corresponds to an FA configuration where the impurity is localized at sites $i=2$ and $i=M-1$, while the second corresponds to an FA configuration where the impurity is localized at $i=3$ and $i=M-2$. We remind the reader that the latter configuration was not observed for $M=12$. Nevertheless, such a configuration only appears for a small range of interactions.

\section{Continuous model}
\label{app:continous}

The employed Hubbard model works for lattices where the tight-binding approximation is valid. However, it is relevant to examine if the presented results hold in shallower lattices. To investigate this, in this appendix, we consider a three-component system in the one-dimensional continuum under the presence of an oscillatory confining potential. This is described by the following Hamiltonian
\begin{align}
    \Hop=&\sum_{\sigma=\uparrow,\downarrow,I}\sum_{i=1}^{N_\sigma}\left[-\frac{\hbar^2}{2m^{\sigma}}\frac{\mathrm{d}^2}{\mathrm{d} (x_{i}^{\sigma})^2}+V_0\sin^2\left(\frac{\pi x_{i}^{\sigma} M }{L}\right)\right]\nonumber \\
    &+g_{ff}\sum_{i=1}^{N_\uparrow}\sum_{j=1}^{N_\downarrow}\delta\left(x_{i}^{\uparrow}-x_{i}^{\downarrow}\right)+g_{fI}\sum_{\sigma=\uparrow,\downarrow}\sum_{i=1}^{N_\sigma}\delta\left(x_{i}^{\sigma}-x_{i}^{I}\right),
\label{app:continuous;eq:H}
\end{align}
where $x_i^\sigma$ gives the position of particle $i$ of species $\sigma$, $m^\sigma$ is the mass of the particles of species $\sigma$, $g_{ff}$ is the strength of the contact $\uparrow$-$\downarrow$ interaction, $g_{fI}$ is the strength of the contact fermion-impurity interaction, $V_0$ gives the amplitude of the lattice potential, and $L$ corresponds to the total length of the system. As in the main text, $M$ is the number of sites. To characterize the depth of the lattice and to connect to experiments, the lattice depth will be rescaled in units of the recoil energy~\cite{lewenstein_ultracold_2012}
\begin{equation}
    E_R=\frac{4\pi^2}{2m\lambda^2},
\end{equation}
where $\lambda=2L/M$ is the wavelength of the lattice. The particles are assumed to have the same mass such that $m^\uparrow = m^\downarrow = m^I = m$. 

In order to solve Hamiltonian~(\ref{app:continuous;eq:H}), it is more convenient to use the exact diagonalization method in the language of second quantization. The Hamiltonian can be rewritten as~\cite{anh-tai_engineering_2024}
\begin{align}
    \Hop = &\sum_{\sigma=\uparrow,\downarrow,I} \sum_{i,j} h^\sigma_{ij}\aop^\dagger_{i,\sigma}\aop_{j,\sigma} \nonumber \\
    &+\sum_{\sigma \neq \gamma \in \{\uparrow,\downarrow,I \}}\sum_{ijk\ell}W^{\sigma\gamma}_{ijk\ell}\aop^\dagger_{i,\sigma}\aop^\dagger_{j,\gamma}\aop_{\ell,\gamma}\aop_{k,\sigma},
    \label{app:continuous;eq:H-2nd}
\end{align}
where
\begin{equation}
    h_{ij}^\sigma = \int \varphi^*_{i,\sigma}(x^\sigma) \Hop_{sp}^\sigma \varphi_{i,\sigma}(x^\sigma) dx^\sigma,
\end{equation}
are the one-body integrals, with $\Hop_{sp}^\sigma = -\frac{\hbar^2}{2m^{\sigma}}\frac{\mathrm{d}^2}{\mathrm{d} (x^{\sigma})^2}+V_0\sin^2\left(\frac{\pi x^{\sigma} M }{L}\right)$ is the single-particle Hamiltonian of $\sigma-$species. Meanwhile, the inter-species two-body interaction integrals are defined as
\begin{align}
    W_{ijk\ell}^{\sigma\gamma} = g_{\sigma\gamma} \iint \varphi^*_{i,\sigma}(x^\sigma)&\varphi^*_{j,\gamma}(x^\gamma) \delta(x^\sigma-x^\gamma) \cdot \nonumber\\
    &\varphi_{k,\sigma}(x^\sigma)\varphi_{\ell,\gamma}(x^\gamma)   dx^\sigma dx^\gamma.
\end{align}
A complete set of single-particle states $|\varphi_{i,\sigma}(x^\sigma)\rangle$ is chosen as the eigenfunctions of the single-particle Hamiltonian $\Hop_{sp}^\sigma$. The many-body ansatz wave function of the three-component mixture can be expanded as a linear combination of the Fock bases as
\begin{equation}
    |\Psi\rangle = \sum_{i_\uparrow = 1}^{\mathcal{D_\uparrow}}\sum_{i_\downarrow = 1}^{\mathcal{D_\downarrow}} \sum_{i_I = 1}^{\mathcal{D}_I} c_{i_\uparrow,i_\downarrow,i_I}|n^\uparrow\rangle_{i_\uparrow}|n^\downarrow\rangle_{i_\downarrow}|n^I\rangle_{i_I},    
\end{equation}
where $c_{i_\uparrow,i_\downarrow,i_I}$ denotes the expansion coefficients, while $|n^\sigma\rangle_{i_\sigma} = |n_1^\sigma,n_2^\sigma,\cdots,n^\sigma_{\mathcal{M_\sigma}} \rangle$ are the Fock configurations of the $\sigma-$component, truncated at the $\mathcal{M}_\sigma$ single-particle mode. One can find the eigen-pairs of the many-body Hamiltonian \eqref{app:continuous;eq:H-2nd} by numerically diagonalizing its matrix representation with respect to the ansatz.

We consider two fermions in each spin component ($N_\uparrow=N_\downarrow=2$) and one impurity ($N_I=1$) in a lattice with $M=8$, resulting in a quarter filling. After the diagonalizations, we extract the density profiles $n_{\sigma}(x)$ ($\sigma=\uparrow,\downarrow,I$), where $x$ is the length coordinate. Numerical convergence was tested by including up to $\mathcal{M}_\sigma = 32$ single-particle modes for each component. We find that stable convergence is achieved when the total number of modes satisfies $\mathcal{M}_\sigma = nM$, where $n \in \mathbb{N}$ indexes the $n$-th band of the single-particle energy spectrum. Other choices of $\mathcal{M_\sigma}$ can lead to numerical instability. In the following, the length is chosen so that $x=0$ is the center of the system.

\begin{figure}[t!]
\centering
\includegraphics[width=\columnwidth]{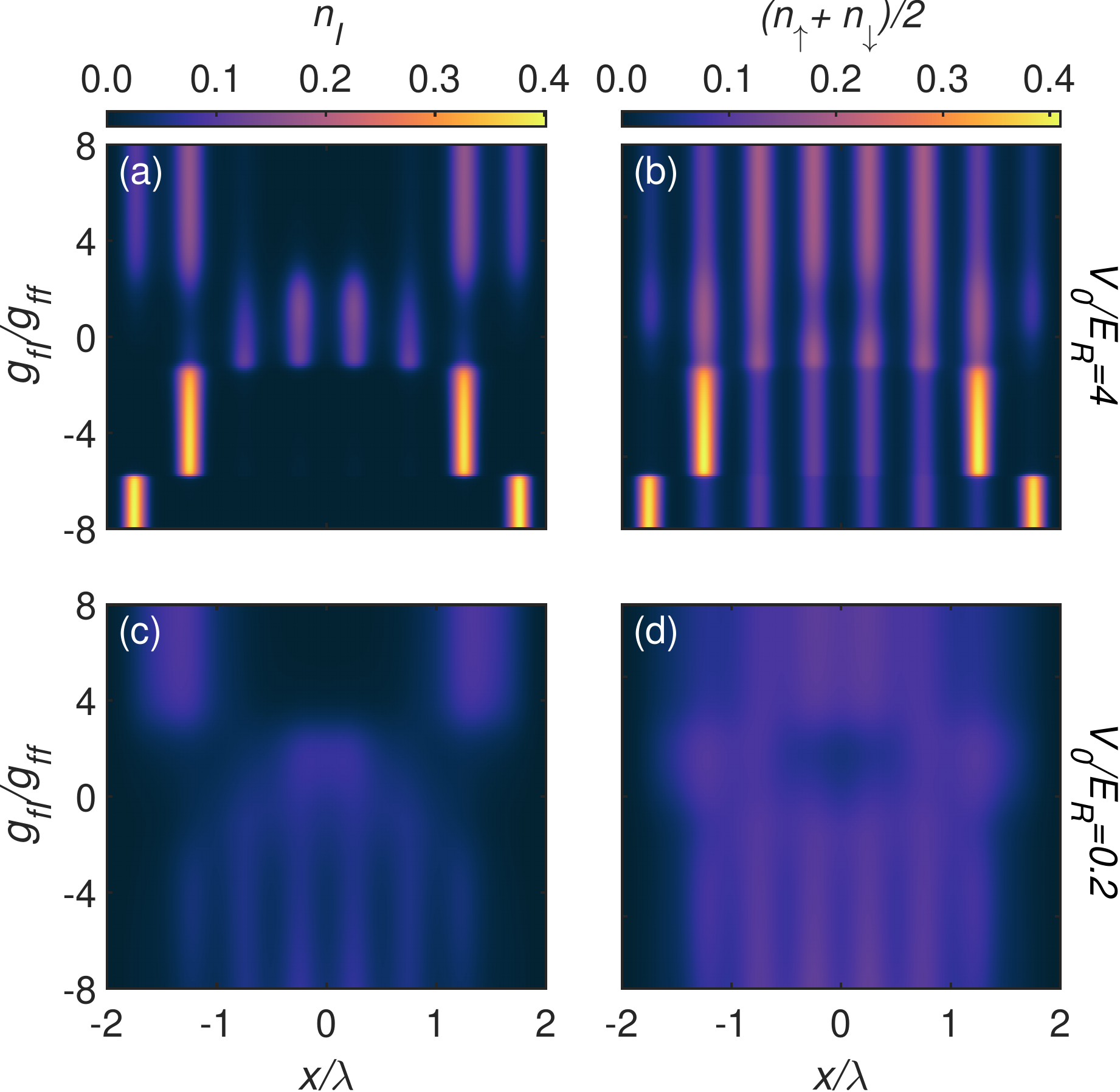}
\caption{Density profiles of the impurity [(a) and (c)] and of the fermions [(b) and (d)]  at $\nu_f=1/4$  as a function of $g_{fI}$ and of the distance $x$ in a lattice with $M=12$. The top and bottom panels consider $V_0/E_R=4$ and $V_0/E_R=0.2$, respectively.}
\label{app:continuous;fig:heatmapsg}
\end{figure}

We show density profiles of the impurity and the fermions across a wide range of fermion-impurity interactions in Fig.~\ref{app:continuous;fig:heatmapsg}, complementing Fig.~\ref{sec:1o4;sub:densities;fig:heatmaps} of the main text. The top panels consider a lattice with a depth much larger than the recoil energy, and thus, it is in the tight-binding regime. These show a remarkable qualitative agreement with the behavior reported in the main text. Indeed, the continuous system exhibits all the regimes previously reported, including both phase-separated configurations and the Friedel attractive regime. In particular, we observe the localization of the impurity at sites $i=2$ and $i=M-1$ for intermediate attraction. 

A shallow lattice is then considered in the bottom panels of Fig.~\ref{app:continuous;fig:heatmapsg}. Here, all the particles show a very weak localization at the sites due to the small depth of the wells. In this case, we do not observe any clear Friedel pattern, showing that the oscillations disappear in shallower lattices. Additionally, and also very importantly, while the pPS persists for strong repulsion, the hPS is not present at strong attraction. This is because the particle-hole symmetry is a property of the Hubbard model, and thus can only be reproduced in tight-binding lattices. In contrast, in shallow lattices, the system does not display this symmetry. Therefore, for small $V_0$, the system only supports a standard phase separation between particles and impurity for strong repulsion, without supporting a phase separation with holes for attractive interactions.

\begin{figure}[t!]
\centering
\includegraphics[width=\columnwidth]{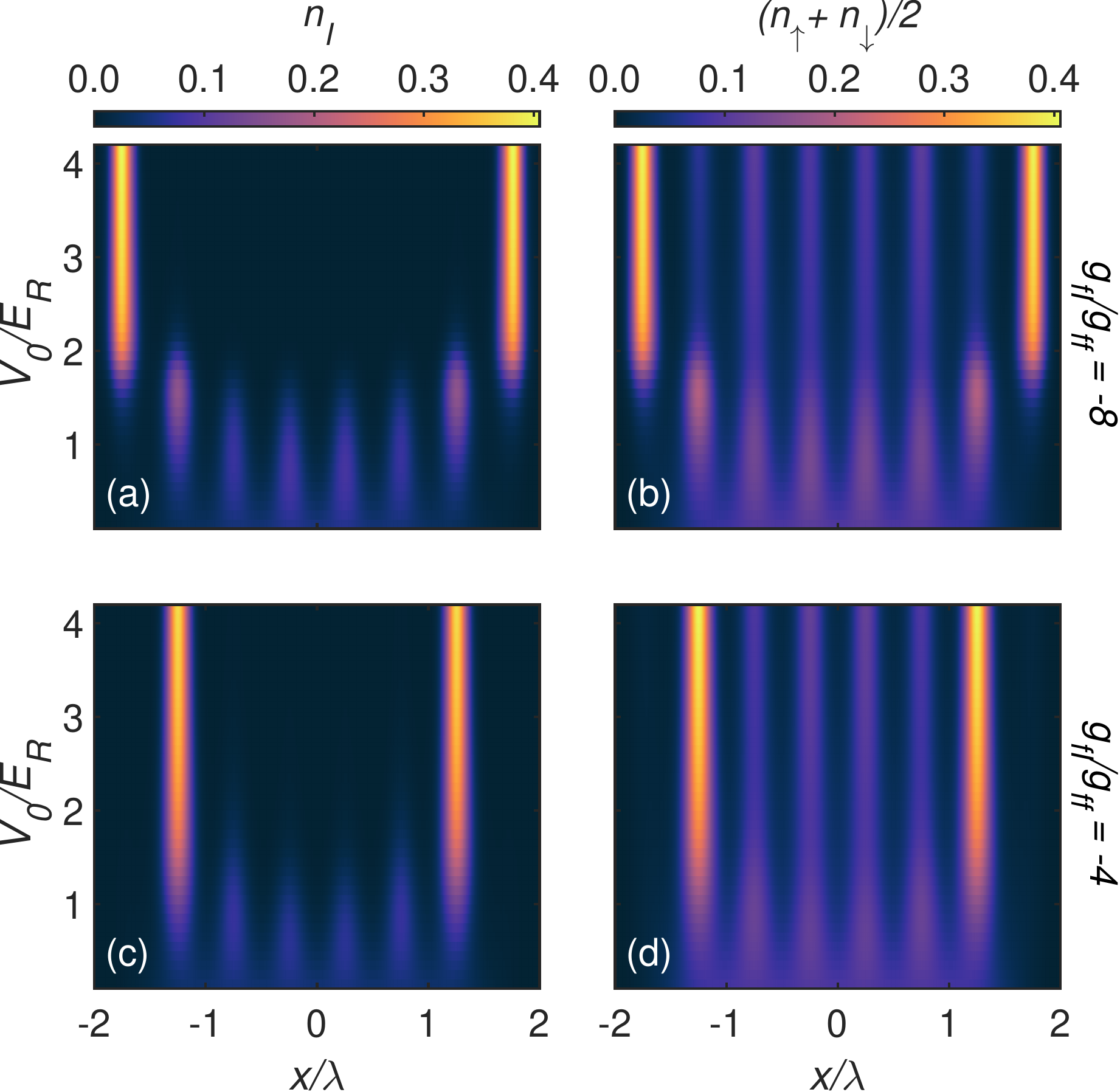}
\caption{Density profiles of the impurity [(a) and (c)] and of the fermions [(b) and (d)]  at $\nu_f=1/4$  as a function of the lattice depth $V_0$ and of the distance $x$ in a lattice with $M=12$ sites. The top and bottom panels consider $g_{fI}/g_{ff}=-8$ and $g_{fI}/g_{ff}=-4$, respectively.}
\label{app:continuous;;fig:heatmapsV0}
\end{figure}

Finally, to find when the hPS and the FA regimes appear, in Fig.~\ref{app:continuous;;fig:heatmapsV0} we show densities as a function of the lattice depth $V_0$. For strong attraction (top panels), no hPS is present for small $V_0$ (bottom part of the panels). However, we observe that the hPS configuration appears for $V_0/E_R\gtrsim 2$, where the impurity sharply localizes at the borders of the lattice. Then, in the bottom panels, we consider an intermediate attraction. Similarly, no Friedel pattern is observed for small $V_0$. However, these also appear for $V_0/E_R\gtrsim 1$, where the impurity localizes at sites $i=2$ and $i=M-1$. Surprisingly, these transitions occur at relatively small lattice depths, for $V_0/E_R\gtrsim 1$.

\newpage

\bibliography{biblio}

\end{document}